\documentclass[12pt]{article}
\usepackage{epsf}
\usepackage{rotate}
\renewcommand{\thefootnote}{\fnsymbol{footnote}}
\newcommand{\lsim}{
\mathrel{\hbox{\rlap{\hbox{\lower4pt\hbox{$\sim$}}}\hbox{$<$}}}}
\newcommand{\gsim}{
\mathrel{\hbox{\rlap{\hbox{\lower4pt\hbox{$\sim$}}}\hbox{$>$}}}}

\begin{document}


\begin{titlepage}
\begin{flushright}
\begin{tabular}{l}
DESY 02--040\\
UAB--FT--523\\
hep--ph/0204101\\
April 2002
\end{tabular}
\end{flushright}

\vspace*{1.3truecm}

\begin{center}
\boldmath
{\Large \bf Exploring CP Violation through Correlations in}

\vspace*{0.3truecm}

{\Large \bf $B\to\pi K$, $B_d\to\pi^+\pi^-$, $B_s\to K^+K^-$
Observable Space}
\unboldmath

\vspace*{1.6cm}

\smallskip
\begin{center}
{\sc {\large Robert Fleischer}}\footnote{E-mail: {\tt
Robert.Fleischer@desy.de}} \\
\vspace*{2mm}
{\sl Deutsches Elektronen-Synchrotron DESY, Notkestra\ss e 85,
D--22607 Hamburg, Germany}
\vspace*{1truecm}\\
{\sc{\large Joaquim Matias}}\footnote{E-mail: {\tt
              matias@ifae.es}}\\
\vspace*{2mm}
{\sl IFAE, Universitat Aut\`onoma de Barcelona, E--08193 Barcelona, Spain}
\end{center}

\vspace{2.0truecm}

{\large\bf Abstract\\[10pt]} \parbox[t]{\textwidth}{
We investigate allowed regions in observable space of $B\to\pi K$,
$B_d\to\pi^+\pi^-$ and $B_s\to K^+K^-$ decays, characterizing these
modes in the Standard Model. After a discussion of a new kind of
contour plots for the $B\to\pi K$ system, we focus on the mixing-induced 
and direct CP asymmetries of the decays $B_d\to\pi^+\pi^-$ and 
$B_s\to K^+K^-$. Using experimental information on the CP-averaged 
$B_d\to\pi^\mp K^\pm$ and $B_d\to\pi^+\pi^-$ branching ratios, the 
relevant hadronic penguin parameters can be constrained, implying 
certain allowed regions in observable space. In the case of 
$B_d\to\pi^+\pi^-$, an interesting situation arises now in view of the
recent $B$-factory measurements of CP violation in this channel, allowing
us to obtain new constraints on the CKM angle $\gamma$ as a function of
the $B^0_d$--$\overline{B^0_d}$ mixing phase $\phi_d=2\beta$, which is 
fixed through ${\cal A}_{\rm CP}^{\rm mix}(B_d\to J/\psi K_{\rm S})$ 
up to a twofold ambiguity. If we assume that 
${\cal A}_{\rm CP}^{\rm mix}(B_d\to \pi^+\pi^-)$ is positive, as indicated
by recent Belle data, and that $\phi_d$ is in agreement with the ``indirect'' 
fits of the unitarity triangle, also the corresponding values for $\gamma$ 
around $60^\circ$ can be accommodated. On the other hand, for the second 
solution of $\phi_d$, we obtain a gap around $\gamma\sim60^\circ$. The allowed 
region in the space of ${\cal A}_{\rm CP}^{\rm mix}(B_s\to K^+K^-)$ and
${\cal A}_{\rm CP}^{\rm dir}(B_s\to K^+K^-)$ is very constrained in the
Standard Model, thereby providing a narrow target range for run II of the
Tevatron and the experiments of the LHC era.
}

\vskip1.5cm

\end{center}

\end{titlepage}

\thispagestyle{empty}
\vbox{}
\newpage

\setcounter{page}{1}

\setcounter{footnote}{0}
\renewcommand{\thefootnote}{\arabic{footnote}}

\section{Introduction}\label{sec:intro}
One of the most exciting aspects of present particle physics is the
exploration of CP violation through $B$-meson decays, allowing us to
overconstrain both the sides and the three angles $\alpha$, $\beta$ and
$\gamma$ of the usual non-squashed unitarity triangle of the
Cabibbo--Kobayashi--Maskawa (CKM) matrix \cite{revs}. Besides the
``gold-plated'' mode $B_d\to J/\psi K_{\rm S}$ \cite{bisa},
which has recently led to the observation of CP violation in the $B$ system
\cite{BaBar-CP-obs,Belle-CP-obs}, there are many different avenues we
may follow to achieve this goal.

In this paper, we first consider $B\to\pi K$ modes
\cite{GRL}--\cite{BF-neutral2}, and then focus on the $B_d\to\pi^+\pi^-$,
$B_s\to K^+K^-$ system \cite{RF-BsKK}, providing promising strategies to
determine $\gamma$. In a previous paper \cite{Fl-Ma}, we pointed out that
these non-leptonic $B$ decays can be characterized
efficiently within the Standard Model through allowed regions in the space
of their observables. If future measurements should result in values for
these quantities lying significantly outside of these regions, we would
have an immediate indication for the presence of new physics. On the other
hand, a measurement of observables lying inside these regions would allow
us to extract values for the angle $\gamma$, which may then show
discrepancies with other determinations, thereby also indicating new physics.
Since penguin processes play a key r\^ole in $B\to\pi K$, $B_d\to\pi^+\pi^-$
and $B_s\to K^+K^-$ decays, these transitions actually represent sensitive
probes for physics beyond the Standard Model \cite{BpiK-NP}.

Besides an update and extended discussion of the allowed regions in 
observable space of appropriate combinations of $B\to\pi K$ decays, 
following \cite{Fl-Ma}, the main point of the present paper is a detailed
analysis of the $B_d\to\pi^+\pi^-$, $B_s\to K^+K^-$ system in the light of
recent experimental data. These neutral $B$-meson decays into final CP
eigenstates provide a time-dependent CP asymmetry of the following form:
\begin{eqnarray}
\lefteqn{a_{\rm CP}(t)\equiv\frac{\Gamma(B^0_q(t)\to f)-
\Gamma(\overline{B^0_q}(t)\to f)}{\Gamma(B^0_q(t)\to f)+
\Gamma(\overline{B^0_q}(t)\to f)}}\nonumber\\
&&=\left[\frac{{\cal A}_{\rm CP}^{\rm dir}(B_q\to f)\,\cos(\Delta M_q t)+
{\cal A}_{\rm CP}^{\rm mix}(B_q\to f)\,\sin(\Delta
M_q t)}{\cosh(\Delta\Gamma_qt/2)-{\cal A}_{\rm
\Delta\Gamma}(B_q\to f)\,\sinh(\Delta\Gamma_qt/2)}
\right],\label{time-CP-asym}
\end{eqnarray}
where we have separated, as usual, the ``direct'' from the ``mixing-induced''
CP-violating contributions. The time-dependent rates refer to initially,
i.e.\ at time $t=0$, present $B^0_q$- or $\overline{B^0_q}$-mesons,
$\Delta M_q>0$ denotes the mass difference of the $B_q$ mass eigenstates,
and $\Delta\Gamma_q$ is their decay width difference, which is negligibly
small in the $B_d$ system, but may be as large as ${\cal O}(10\%)$ in the
$B_s$ system \cite{DG-Bs}. The three observables in (\ref{time-CP-asym})
are not independent from one another, but satisfy the
following relation:
\begin{equation}
\Bigl[{\cal A}_{\rm CP}^{\rm dir}(B_q\to f)\Bigr]^2+
\Bigl[{\cal A}_{\rm CP}^{\rm mix}(B_q\to f)\Bigr]^2
+\Bigl[{\cal A}_{\rm \Delta\Gamma}(B_q\to f)\Bigr]^2=1.
\end{equation}

If we employ the $U$-spin flavour symmetry of strong interactions,
relating down and strange quarks to each other, the CP-violating observables
provided by $B_d\to\pi^+\pi^-$ and $B_s\to K^+K^-$ allow a determination
both of $\gamma$ and of the $B^0_d$--$\overline{B^0_d}$ mixing phase 
$\phi_d$, which is given by $2\beta$ in the Standard Model \cite{RF-BsKK}. 
Moreover, interesting hadronic penguin parameters can be extracted as well,
consisting of a CP-conserving strong phase, and a ratio of strong amplitudes,
measuring -- sloppily speaking -- the ratio of penguin- to tree-diagram-like
contributions to $B_d\to\pi^+\pi^-$.  
The use of $U$-spin arguments in this approach can be minimized, if we
use $\phi_d$ as an input. As is well known, this phase can be determined from 
mixing-induced CP violation in $B_d\to J/\psi K_{\rm S}$,
\begin{equation}\label{ACP-BpsiK}
{\cal A}_{\rm CP}^{\rm mix}(B_d\to J/\psi K_{\rm S})=-\sin\phi_d,
\end{equation}
up to a twofold ambiguity. Using the present world average
\begin{equation}\label{ACD-BdpsiKS}
\sin\phi_d=0.78\pm0.08,
\end{equation}
which takes into account the most recent results by BaBar \cite{BABAR-new}
and Belle \cite{BELLE-new}, as well as previous results by CDF \cite{CDF}
and ALEPH \cite{ALEPH}, we obtain
\begin{equation}\label{phid-det}
\phi_d=\left(51^{+8}_{-7}\right)^\circ \, \lor \,
\left(129^{+7}_{-8}\right)^\circ.
\end{equation}
On the other hand, the $B^0_s$--$\overline{B^0_s}$ mixing phase $\phi_s$,
which enters ${\cal A}_{\rm CP}^{\rm mix}(B_s\to K^+K^-)$,  is
negligibly small in the Standard Model. It should be noted that we have 
assumed in (\ref{ACP-BpsiK}) that new-physics contributions to the
$B\to J/\psi K$ decay amplitudes are negligible. This assumption can be
checked through the observable set introduced in \cite{FM-BpsiK}.

Whereas $B_d\to\pi^+\pi^-$ is already accessible at the $e^+e^-$
$B$-factories operating at the $\Upsilon(4S)$ resonance, BaBar, Belle and
CLEO, the $B_s\to K^+K^-$ mode can be studied nicely at hadron machines,
i.e.\ at run II of the Tevatron and at the experiments of the LHC era,
where the strategy sketched above may lead to experimental accuracies for 
$\gamma$ of ${\cal O}(10^\circ)$ \cite{CDF-2} and ${\cal O}(1^\circ)$ 
\cite{LHC-Report}, respectively. Unfortunately, experimental data on 
$B_s\to K^+K^-$ are not yet available. However, since $B_s\to K^+K^-$ is 
related to $B_d\to\pi^\mp K^\pm$ through an interchange of spectator quarks, 
$SU(3)$ flavour-symmetry arguments and plausible dynamical assumptions allow 
us to replace $B_s\to K^+K^-$ approximately by $B_d\to\pi^\mp K^\pm$,
which can already be explored at the $B$-factories. A key element of our 
analysis is the ratio of the CP-averaged $B_d\to\pi^+\pi^-$ and 
$B_d\to\pi^\mp K^\pm$ branching ratios, which can be expressed in terms of 
$\gamma$ and hadronic penguin parameters. As pointed out in 
\cite{RF-pen-constr}, constraints on the latter quantities can be obtained 
from this observable, allowing an interesting comparison with theoretical
predictions. 

In our analysis, we shall follow these lines to explore also allowed 
regions in the space of the CP asymmetries of the $B_d\to\pi^+\pi^-$, 
$B_s\to K^+K^-$ system, and constraints on $\gamma$. To this end, we first 
use (\ref{ACP-BpsiK}) to fix the $B^0_d$--$\overline{B^0_d}$ mixing phase 
$\phi_d$, yielding the twofold solution (\ref{phid-det}). For a given value 
of the mixing-induced CP asymmetry ${\cal A}_{\rm CP}^{\rm mix}(B_d\to
\pi^+\pi^-)$, the ratio of the CP-averaged $B_d\to\pi^+\pi^-$ and 
$B_d\to\pi^\mp K^\pm$ branching ratios allows us then to determine the direct 
CP asymmetry ${\cal A}_{\rm CP}^{\rm dir}(B_d\to\pi^+\pi^-)$ as a function of 
$\gamma$. Consequently, measuring these observables, we may extract this 
angle. Moreover, the corresponding hadronic penguin parameters can be 
determined as well. On the other hand, if we assume that 
${\cal A}_{\rm CP}^{\rm mix}(B_d\to\pi^+\pi^-)$ lies within a certain given 
range, bounds on ${\cal A}_{\rm CP}^{\rm dir}(B_d\to\pi^+\pi^-)$ and $\gamma$ 
can be obtained, depending on the choice of $\phi_d$. In particular, 
we may assume that the mixing-induced CP asymmetry 
${\cal A}_{\rm CP}^{\rm mix}(B_d\to\pi^+\pi^-)$ is positive or negative,
leading to very different situations. 

Since experimental data for the direct and mixing-induced CP 
asymmetries of $B_d\to\pi^+\pi^-$ are already available from the $B$ 
factories, we may now start to fill these strategies with 
life:\footnote{The connection between our notation and those
employed in \cite{BaBar-Bpipi-new,Belle-Bpipi} is as follows:
${\cal A}_{\rm CP}^{\rm dir}(B_d\to\pi^+\pi^-)=+C_{\pi\pi}^{\rm BaBar}=
-{\cal A}_{\pi\pi}^{\rm Belle}$ and 
${\cal A}_{\rm CP}^{\rm mix}(B_d\to\pi^+\pi^-)
=-S_{\pi\pi}^{\rm BaBar}=-S_{\pi\pi}^{\rm Belle}$.}
\begin{equation}\label{Adir-exp}
{\cal A}_{\rm CP}^{\rm dir}(B_d\to\pi^+\pi^-)=\left\{
\begin{array}{ll}
-0.02\pm0.29\pm0.07 & \mbox{(BaBar \cite{BaBar-Bpipi-new})}\\
-0.94^{+0.31}_{-0.25}\pm0.09 & \mbox{(Belle \cite{Belle-Bpipi})}
\end{array}
\right.
\end{equation}
\begin{equation}\label{Amix-exp}
{\cal A}_{\rm CP}^{\rm mix}(B_d\to\pi^+\pi^-)=\left\{
\begin{array}{ll}
0.01\pm0.37\pm0.07& \mbox{(BaBar \cite{BaBar-Bpipi-new})}\\
1.21^{+0.27+0.13}_{-0.38-0.16} & \mbox{(Belle \cite{Belle-Bpipi}),}
\end{array}
\right.
\end{equation}
yielding the na\"\i ve averages 
\begin{equation}\label{CP-Bpipi-average}
{\cal A}_{\rm CP}^{\rm dir}(B_d\to\pi^+\pi^-)=-0.48\pm0.21, \quad
{\cal A}_{\rm CP}^{\rm mix}(B_d\to\pi^+\pi^-)=0.61\pm0.26.
\end{equation}
Unfortunately, the BaBar results, which are an update of the values given 
in \cite{BaBar-Bpipi}, and those of the first Belle measurement are
not fully consistent with one another. In contrast to BaBar, Belle signals 
large direct and mixing-induced CP violation in $B_d\to\pi^+\pi^-$, and 
points towards a positive value of ${\cal A}_{\rm CP}^{\rm mix}(B_d\to
\pi^+\pi^-)$. As we shall point out in this paper, the following picture
arises now: for a positive observable ${\cal A}_{\rm CP}^{\rm mix}
(B_d\to \pi^+\pi^-)$, as indicated by Belle, the solution of $\phi_d$ being 
in agreement with the ``indirect'' fits of the unitarity 
triangle~\cite{UT-fits}, yielding $\phi_d\sim 45^\circ$, 
allows us to accommodate also the corresponding values for $\gamma$ around 
$60^\circ$, whereas a gap around $\gamma\sim60^\circ$ arises for the second 
solution of $\phi_d$. On the other hand, varying ${\cal A}_{\rm CP}^{\rm mix}
(B_d\to\pi^+\pi^-)$ within its whole negative range, $\gamma$ remains rather 
unconstrained in the physically most interesting region. Using the 
experimental averages given in (\ref{CP-Bpipi-average}), we obtain 
$28^\circ\lsim\gamma\lsim74^\circ$ $(\phi_d=51^\circ)$ and 
$106^\circ\lsim\gamma\lsim152^\circ$ $(\phi_d=129^\circ)$. Interestingly,
there are some indications that $\gamma$ may actually be larger than 
$90^\circ$, which may then point towards the unconventional solution of 
$\phi_d=129^\circ$. The negative sign of ${\cal A}_{\rm CP}^{\rm dir}
(B_d\to\pi^+\pi^-)$ implies that a certain CP-conserving strong phase 
$\theta$ has to lie within the range $0^\circ<\theta<180^\circ$. In the 
future, improved experimental data will allow us to extract $\gamma$ and 
the relevant hadronic parameters in a much more stringent way 
\cite{RF-BsKK,RF-pen-constr}. 

Following a different avenue, implications of the measurements of the CP 
asymmetries of $B_d\to\pi^+\pi^-$ were also investigated by Gronau and Rosner 
in \cite{GR-Bpipi}. The main differences to our analysis are as follows: in 
\cite{GR-Bpipi}, the $B_d\to\pi^+\pi^-$ observables are expressed in terms of 
$\alpha$ and $\beta$, the ``tree'' amplitude $T_{\pi\pi}$ is estimated using 
factorization and data on $B\to\pi \ell\nu$, and the ``penguin'' amplitude 
$P_{\pi\pi}$ is fixed through the CP-averaged $B^\pm\to\pi^\pm K$ 
branching ratio with the help of $SU(3)$ flavour-symmetry and plausible 
dynamical assumptions. In contrast, we express the observables in terms of 
$\gamma$ and the general $B^0_d$--$\overline{B^0_d}$ mixing phase $\phi_d$, 
which is equal to $2\beta$ in the Standard Model, and use the ratio of the 
CP-averaged $B_d\to\pi^+\pi^-$ and $B_d\to\pi^\mp K^\pm$ branching ratios as 
an additional observable to deal with the penguin contributions, requiring 
also $SU(3)$ flavour-symmetry and plausible dynamical assumptions. We prefer 
to follow these lines, since we have then not to make a separation between 
tree and penguin amplitudes, which is complicated by long-distance 
contributions, and have not to use factorization to estimate the 
overall magnitude of the tree-diagram-dominated amplitude $T_{\pi\pi}$; 
factorization is only used in our approach to take into account 
$SU(3)$-breaking effects. As far as the weak phases are concerned, we 
prefer to use $\gamma$ and $\phi_d$, since the 
results for the former quantity can then be compared directly with 
constraints from other processes, whereas the latter can anyway be determined 
straighforwardly from mixing-induced CP violation in $B_d\to J/\psi K_{\rm S}$
up to a twofold ambiguity, also if there should be CP-violating new-physics 
contributions to $B^0_d$--$\overline{B^0_d}$ mixing. This way, we obtain an 
interesting link between the two solutions for $\phi_d$ and the allowed 
ranges for $\gamma$, as we have noted above. 

It should be emphasized that the 
parametrization of the CP-violating $B_d\to\pi^+\pi^-$ observables in terms 
of $\gamma$ and $\phi_d$ is actually more direct than the one in terms of 
$\alpha$ and $\beta$, as the appearance of $\alpha$ is due to the elimination
of $\gamma$ with the help of the unitarity relation $\gamma=180^\circ-
\alpha-\beta$. If there were negligible penguin contributions to 
$B_d\to\pi^+\pi^-$, mixing-induced CP violation in this channel would allow 
us to determine the combination $\phi_d+2\gamma$, which is equal to 
$-2\alpha$ in the Standard Model. On the other hand, in the presence of 
significant penguin contributions, as indicated by experimental data, it is 
actually more advantageous to keep $\phi_d$ and $\gamma$ in the 
parametrization of the $B_d\to\pi^+\pi^-$ observables. Moreover, we may 
then also investigate straightforwardly the impact of possible CP-violating 
new-physics contributions to $B^0_d$--$\overline{B^0_d}$ mixing, which may
yield the unconventional value of $\phi_d=129^\circ$. These features will 
become obvious when we turn to the details of our approach. 

Another important aspect of our study is an analysis of the decay 
$B_s\to K^+K^-$, which is particularly promising for hadronic $B$ 
experiments. Using the experimental results for the ratio of the CP-averaged 
$B_d\to\pi^+\pi^-$ and $B_d\to\pi^\mp K^\pm$ branching ratios, we obtain a 
very constrained allowed region in the ${\cal A}_{\rm CP}^{\rm mix}
(B_s\to K^+K^-)$--${\cal A}_{\rm CP}^{\rm dir}(B_s\to K^+K^-)$ plane within 
the Standard Model. If future measurements should actually fall into this 
very restricted target range in observable space, the combination of 
$B_s\to K^+K^-$ with $B_d\to\pi^+\pi^-$ through the $U$-spin flavour symmetry 
of strong interactions allows a determination of $\gamma$, as we have noted 
above. On the other hand, if the experimental results should show a 
significant deviation from the Standard-Model range in observable space, a 
very exciting situation would arise immediately, pointing towards new physics.

The outline of this paper is as follows: in Section~\ref{sec:BpiK},
we first turn to the allowed regions in observable space of $B\to\pi K$
decays, and give a new kind of contour plots, allowing us to read off 
directly the preferred ranges for $\gamma$ and strong phases from the
experimental data. In Section~\ref{sec:Bdpipi-BsKK}, we then discuss the 
general formalism to deal with the $B_d\to\pi^+\pi^-$, $B_s\to K^+K^-$
system, and show how constraints on the relevant penguin parameters
can be obtained from data on $B_d\to\pi^\mp K^\pm$. The implications
for the allowed regions in observable space for the decays $B_d\to\pi^+\pi^-$
and $B_s\to K^+K^-$ will be explored in Sections~\ref{sec:Bdpipi} and
\ref{sec:BsKK}, respectively. In our analysis, we shall also discuss the 
impact of theoretical uncertainties, and comment on certain simplifications,
which could be made by using a rather moderate input from factorization.
Finally, we summarize our conclusions and give a brief outlook in 
Section~\ref{sec:concl}.

\boldmath
\section{Allowed Regions in $B\to\pi K$ Observable Space}\label{sec:BpiK}
\unboldmath
\subsection{Amplitude Parametrizations and Observables}
The starting point of analyses of the $B\to\pi K$ system is the
isospin flavour symmetry of strong interactions, which implies the
following amplitude relations:
\begin{displaymath}
\sqrt{2}A(B^+\to\pi^0K^+)+A(B^+\to\pi^+K^0)=
\sqrt{2}A(B^0_d\to\pi^0K^0)+A(B^0_d\to\pi^-K^+)
\end{displaymath}
\begin{equation}\label{ampl-rel}
=-\biggl[|T+C|e^{i\delta_{T+C}}e^{i\gamma}+
P_{\rm ew}\biggr]\propto\left[e^{i\gamma}+q_{\rm ew}\right].
\end{equation}
Here $T$ and $C$ denote the strong amplitudes describing colour-allowed
and colour-suppressed tree-diagram-like topologies, respectively,
$P_{\rm ew}$ is due to colour-allowed and colour-suppressed EW penguins,
$\delta_{T+C}$ is a CP-conserving strong phase, and $q_{\rm ew}$ denotes
the ratio of EW to tree-diagram-like topologies. A relation with an
analogous phase structure holds also for the ``mixed'' $B^+\to\pi^+ K^0$,
$B_d^0\to\pi^- K^+$ system. Because of these relations, the following
combinations of $B\to\pi K$ decays were considered in the literature to
probe $\gamma$:
\begin{itemize}
\item The ``mixed'' $B^\pm\to\pi^\pm K$, $B_d\to\pi^\mp K^\pm$
system \cite{PAPIII}--\cite{defan}.
\item The ``charged'' $B^\pm\to\pi^\pm K$, $B^\pm\to\pi^0K^\pm$
system \cite{NR}--\cite{BF-neutral1}.
\item The ``neutral'' $B_d\to\pi^0 K$, $B_d\to\pi^\mp K^\pm$
system \cite{BF-neutral1,BF-neutral2}.
\end{itemize}
Interestingly, already CP-averaged $B\to\pi K$ branching ratios may lead
to non-trivial constraints on $\gamma$ \cite{FM,NR}. In order to go
beyond these bounds and to determine $\gamma$, also CP-violating rate
differences have to be measured. To this end, it is convenient to introduce
the following sets of observables \cite{BF-neutral1}:
\begin{equation}\label{mixed-obs}
\left\{\begin{array}{c}R\\A_0\end{array}\right\}
\equiv\left[\frac{\mbox{BR}(B^0_d\to\pi^-K^+)\pm
\mbox{BR}(\overline{B^0_d}\to\pi^+K^-)}{\mbox{BR}(B^+\to\pi^+K^0)+
\mbox{BR}(B^-\to\pi^-\overline{K^0})}\right]\frac{\tau_{B^+}}{\tau_{B^0_d}}
\end{equation}
\begin{equation}\label{charged-obs}
\left\{\begin{array}{c}R_{\rm c}\\A_0^{\rm c}\end{array}\right\}
\equiv2\left[\frac{\mbox{BR}(B^+\to\pi^0K^+)\pm
\mbox{BR}(B^-\to\pi^0K^-)}{\mbox{BR}(B^+\to\pi^+K^0)+
\mbox{BR}(B^-\to\pi^-\overline{K^0})}\right]
\end{equation}
\begin{equation}\label{neut-obs}
\left\{\begin{array}{c}R_{\rm n}\\A_0^{\rm n}\end{array}\right\}
\equiv\frac{1}{2}\left[\frac{\mbox{BR}(B^0_d\to\pi^-K^+)\pm
\mbox{BR}(\overline{B^0_d}\to\pi^+K^-)}{\mbox{BR}(B^0_d\to\pi^0K^0)+
\mbox{BR}(\overline{B^0_d}\to\pi^0\overline{K^0})}\right],
\end{equation}
where the $R_{\rm (c,n)}$ are ratios of CP-averaged branching ratios and
the $A_0^{\rm (c,n)}$ represent CP-violating observables. In
Tables~\ref{tab:BPIK-obs} and \ref{tab:BPIK-obs-CPV}, we have summarized
the present status of these quantities implied by the $B$-factory data.
The averages given in these tables were calculated by simply adding the 
errors in quadrature.

\begin{table}[t]
\begin{center}
\begin{tabular}{|c|c|c|c|c|}
\hline
Observable & CLEO \cite{CLEO-BpiK} & BaBar \cite{babar-BpiK} & Belle
\cite{belle-BpiK} & Average\\
\hline
$R$ & $1.00\pm0.30$ & $0.97\pm0.23$ & $1.50\pm0.66$ & $1.16\pm0.25$\\
$R_{\rm c}$ & $1.27\pm0.47$ & $1.19\pm0.35$ & $2.38\pm1.12$ & $1.61\pm0.42$\\
$R_{\rm n}$ & $0.59\pm0.27$ & $1.02\pm0.40$ & $0.60\pm0.29$ & $0.74\pm0.19$\\
\hline
\end{tabular}
\caption{CP-conserving $B\to \pi K$ observables as defined in
(\ref{mixed-obs})--(\ref{neut-obs}). For the evaluation of $R$, we have
used $\tau_{B^+}/\tau_{B^0_d}=1.060\pm0.029$.}\label{tab:BPIK-obs}
\end{center}
\end{table}

\begin{table}[t]
\begin{center}
\begin{tabular}{|c|c|c|c|c|}
\hline
Observable & CLEO \cite{CLEO-BpiK-CPV} & BaBar 
\cite{BaBar-Bpipi-new,babar-BpiK}
& Belle \cite{Belle-Bpipi,belle-BpiK-CPV} & Average\\
\hline
$A_0$ & $0.04\pm0.16$ & $0.05\pm0.06$ & $0.09\pm0.13$ & $0.06\pm0.07$\\
$A_0^{\rm c}$ & $0.37\pm0.32$ & $0.00\pm0.16$ & $0.14\pm0.51$ & $0.17\pm0.21$\\
$A_0^{\rm n}$ & $0.02\pm0.10$ & $0.05\pm0.07$ & $0.04\pm0.05$ & 
$0.04\pm0.04$\\
\hline
\end{tabular}
\caption{CP-violating $B\to \pi K$ observables as defined in
(\ref{mixed-obs})--(\ref{neut-obs}). For the evaluation
of $A_0$, we have used
$\tau_{B^+}/\tau_{B^0_d}=1.060\pm0.029$.}\label{tab:BPIK-obs-CPV}
\end{center}
\end{table}

The purpose of the following considerations is not the extraction
of $\gamma$, which has been discussed at length in
\cite{PAPIII}--\cite{BF-neutral2}, but an analysis of the allowed
regions in the $R_{\rm (c,n)}$--$A_0^{\rm (c,n)}$ planes arising within
the Standard Model. Here we go beyond our previous paper \cite{Fl-Ma}
in two respects: first, we consider not only the mixed and charged
$B\to\pi K$ systems, but also the neutral one, as advocated in 
\cite{BF-neutral1,BF-neutral2}. Second, we include contours in the 
allowed regions that correspond to given values of $\gamma$ and 
$\delta_{\rm (c,n)}$, thereby allowing us to read off directly the preferred 
ranges for these parameters from the experimental data. The ``indirect'' 
fits of the unitarity triangle favour the range
\begin{equation}\label{gamma-SM}
50^\circ\lsim\gamma\lsim70^\circ,
\end{equation}
which corresponds to the Standard-Model expectation for this angle
\cite{UT-fits}. Since the CP-violating parameter $\varepsilon_K$, describing
indirect CP violation in the neutral kaon system, implies a positive value
of the Wolfenstein parameter $\eta$ \cite{wolf},\footnote{For a negative 
bag parameter $B_K$, which appears unlikely to us, negative $\eta$ 
would be implied \cite{GKN}.} we shall restrict $\gamma$ to 
$0^\circ\leq\gamma\leq 180^\circ$.

\begin{table}[t]
\begin{center}
\begin{tabular}{|c|c|c|c|c|}
\hline
Parameter & CLEO \cite{CLEO-BpiK} & BaBar \cite{babar-BpiK} & Belle
\cite{belle-BpiK} & Average\\
\hline
$r_{\rm c}$ & $0.21\pm0.06$ & $0.21\pm0.05$ & $0.30\pm0.09$ & $0.24\pm0.04$\\
$r_{\rm n}$ & $0.17\pm0.06$ & $0.21\pm0.06$ & $0.19\pm0.12$ & $0.19\pm0.05$\\
\hline
\end{tabular}
\caption{Experimental results for $r_{\rm c}$ and $r_{\rm n}$.}\label{tab:r}
\end{center}
\end{table}

To simplify our analysis, we assume that certain rescattering effects
\cite{res} play a minor r\^ole. Employing the formalism discussed in
\cite{BF-neutral1} (for an alternative description, see \cite{neubert}),
it would be possible to take into account also these effects if they should
turn out to be important. However, both the presently available experimental
upper bounds on $B\to KK$ branching ratios and the recent theoretical
progress due to the development of the QCD factorization approach
\cite{QCD-fact1,QCD-fact2} are not in favour of large rescattering effects.

Following these lines, we obtain for the charged and neutral $B\to\pi K$
systems
\begin{eqnarray}
R_{\rm c,n}&=&1-2r_{\rm c,n}\left(\cos\gamma-q\right)\cos\delta_{\rm c,n}
+v^2r_{\rm c,n}^2\label{R-expr}\\
A_0^{\rm c,n}&=&2r_{\rm c,n}\sin\delta_{\rm c,n}\sin\gamma,\label{A0-expr}
\end{eqnarray}
where $\delta_{\rm c,n}$ denotes a CP-conserving strong phase difference
between tree-diagram-like and penguin topologies, $r_{\rm c,n}$ measures
the ratio of tree-diagram-like to penguin topologies, $q$ corresponds to
the electroweak penguin parameter appearing in (\ref{ampl-rel}), and
\begin{equation}
v=\sqrt{1-2q\cos\gamma+q^2}.
\end{equation}
A detailed discussion of these parametrizations can be found in 
\cite{BF-neutral1}. Using the $SU(3)$ flavour symmetry to fix $|T+C|$ 
through $B^+\to\pi^+\pi^0$ \cite{GRL}, we arrive at
\begin{equation}\label{rc-det}
r_{\rm c}=\sqrt{2}\left|\frac{V_{us}}{V_{ud}}\right|\frac{f_K}{f_\pi}
\frac{|A(B^+\to\pi^+\pi^0)|}{\sqrt{\langle|A(B^\pm\to\pi^\pm K)|^2\rangle}}
\end{equation}
\begin{equation}\label{rn-det}
r_{\rm n}=\left|\frac{V_{us}}{V_{ud}}\right|\frac{f_K}{f_\pi}
\frac{|A(B^+\to\pi^+\pi^0)|}{\sqrt{\langle|A(B_d\to\pi^0 K)|^2\rangle}},
\end{equation}
where the ratio $f_K/f_{\pi}$ of the kaon and pion decay constants takes
into account factorizable $SU(3)$-breaking corrections. In \cite{QCD-fact2},
also non-factorizable effects were investigated and found to play a
minor r\^ole. In Table~\ref{tab:r}, we collect the present experimental
results for $r_{\rm c}$ and $r_{\rm n}$ following from (\ref{rc-det}) and
(\ref{rn-det}), respectively. The electroweak penguin parameter $q$ can be
fixed through the $SU(3)$ flavour symmtery \cite{NR}
(see also \cite{PAPIII}), yielding
\begin{equation}
q=0.71\times\left[\frac{0.38}{R_b}\right],
\end{equation}
with
\begin{equation}\label{Rb-def}
R_b=\left(1-\frac{\lambda^2}{2}\right)\frac{1}{\lambda}\left|
\frac{V_{ub}}{V_{cb}}\right|=0.38\pm0.08.
\end{equation}
Taking into account factorizable $SU(3)$ breaking, the central value of
$0.71$ is shifted to $0.68$. For a detailed analysis within the QCD
factorization approach, we refer the reader to \cite{QCD-fact2}.

We may now use (\ref{R-expr}) to eliminate $\sin\delta_{\rm c,n}$ in
(\ref{A0-expr}):
\begin{equation}\label{A0-constr}
A_0^{\rm c,n}=\pm2r_{\rm c,n}
\sqrt{1-\left[\frac{1-R_{\rm c,n}+v^2r_{\rm c,n}^2}{2r_{\rm c,n}
\left(\cos\gamma-q\right)}\right]^2}\sin\gamma,
\end{equation}
allowing us to calculate $A_0^{\rm c,n}$ for given $R_{\rm c,n}$ as a 
function of $\gamma$; if we vary $\gamma$ between $0^\circ$ and $180^\circ$, 
we obtain an allowed region in the $R_{\rm c,n}$--$A_0^{\rm c,n}$ plane. 
This range can also be obtained by varying $\gamma$ and $\delta_{\rm c,n}$
directly in (\ref{R-expr}) and (\ref{A0-expr}), with 
$0^\circ\leq\gamma\leq180^\circ$ and 
$-180^\circ\leq\delta_{\rm c,n}\leq+180^\circ$.

\begin{figure}[t]
\vspace*{-0.8cm}
$$\hspace*{-1.cm}
\epsfysize=0.2\textheight
\epsfxsize=0.3\textheight
\epsffile{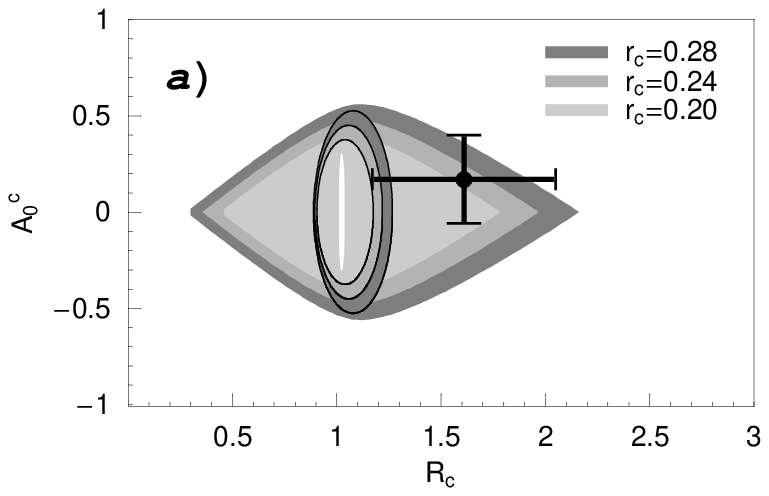} \hspace*{0.3cm}
\epsfysize=0.2\textheight
\epsfxsize=0.3\textheight
\epsffile{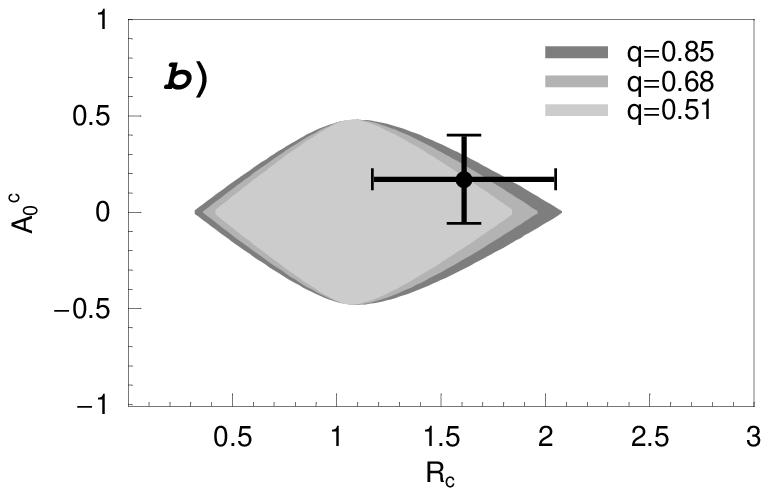}
$$
\vspace*{-0.2cm}
$$\hspace*{-1.cm}
\epsfysize=0.2\textheight
\epsfxsize=0.3\textheight
\epsffile{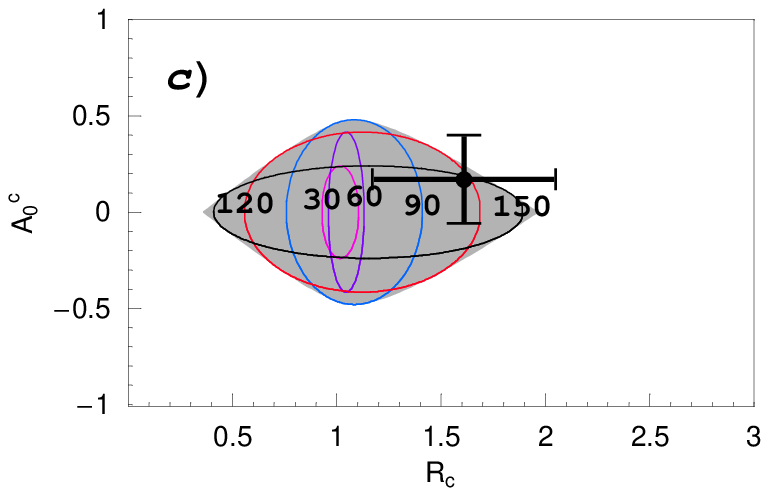} \hspace*{0.3cm}
\epsfysize=0.2\textheight
\epsfxsize=0.3\textheight
\epsffile{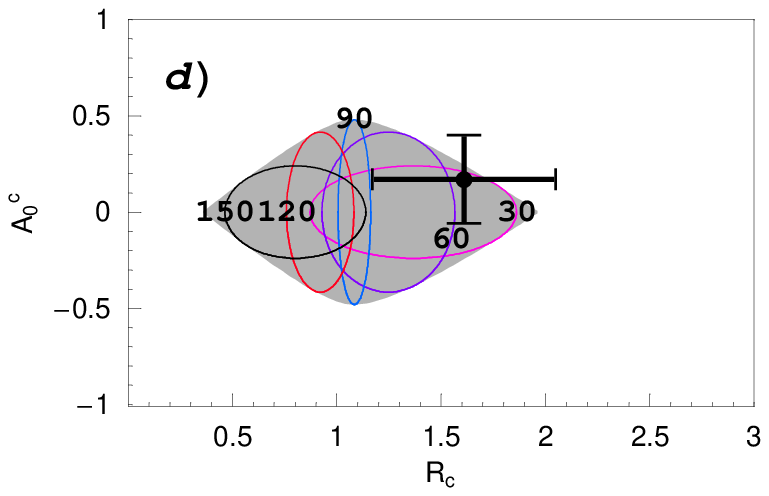}
$$
\vspace*{-0.9cm}
\caption[]{Allowed regions in the $R_{\rm c}$--$A_0^{\rm c}$ plane: (a) 
corresponds to $0.20 \leq r_{\rm c} \leq 0.28$ 
for $q=0.68$, and (b) to $0.51 \leq q \leq 0.85$ for $r_{\rm c}=0.24$;
the elliptical regions arise if we restrict $\gamma$ to the Standard-Model
range (\ref{gamma-SM}). In (c) and (d), we show the contours for fixed 
values of $\gamma$ and $|\delta_{\rm c}|$, respectively 
($r_{\rm c}=0.24$, $q=0.68$).}\label{fig:BpiK-charged}
\end{figure}

A similar exercise can also be performed for the mixed $B\to\pi K$ system.
To this end, we have just to make appropriate replacements of variables in 
(\ref{R-expr}) and (\ref{A0-expr}). Since electroweak penguins contribute 
only in colour-suppressed form to the corresponding decays, we may use 
$q\to0$ in this case to a good approximation. Moreover, we have 
$r_{\rm c,n}\to r$, where the determination of $r$ requires the use of 
arguments related to factorization \cite{PAPIII,GR} to fix the 
colour-allowed amplitude $|T|$, or the measurement of 
$B_s\to\pi^\pm K^\mp$ \cite{GR-U-spin}, which is related to 
$B_d\to\pi^\mp K^\pm$ through the $U$-spin flavour symmetry
of strong interactions. The presently most refined theoretical study of
$r$ can be found in \cite{QCD-fact2}, using the QCD factorization approach.
In our analysis, we shall consider the range $0.14\leq r\leq 0.26$.
Since we have to make use of dynamical arguments to fix $q$ and $r$ in
the case of the mixed $B\to\pi K$ system, it is not as clean as the
charged and neutral $B\to\pi K$ systems.

\begin{figure}
\vspace*{-0.8cm}
$$\hspace*{-1.cm}
\epsfysize=0.2\textheight
\epsfxsize=0.3\textheight
\epsffile{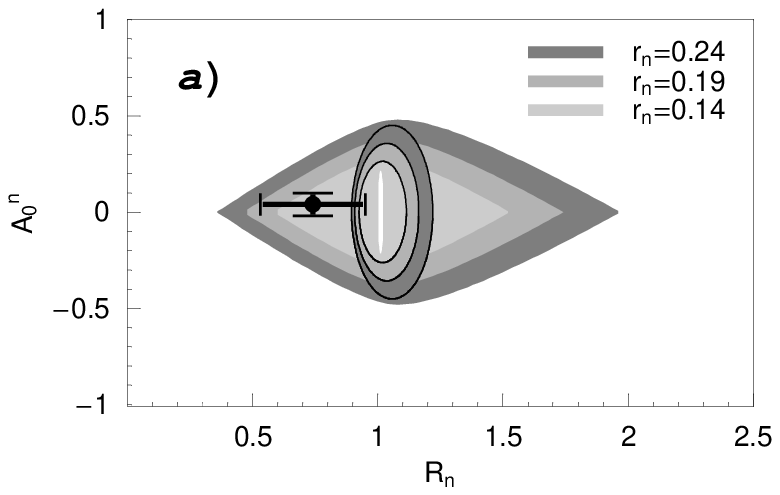} \hspace*{0.3cm}
\epsfysize=0.2\textheight
\epsfxsize=0.3\textheight
\epsffile{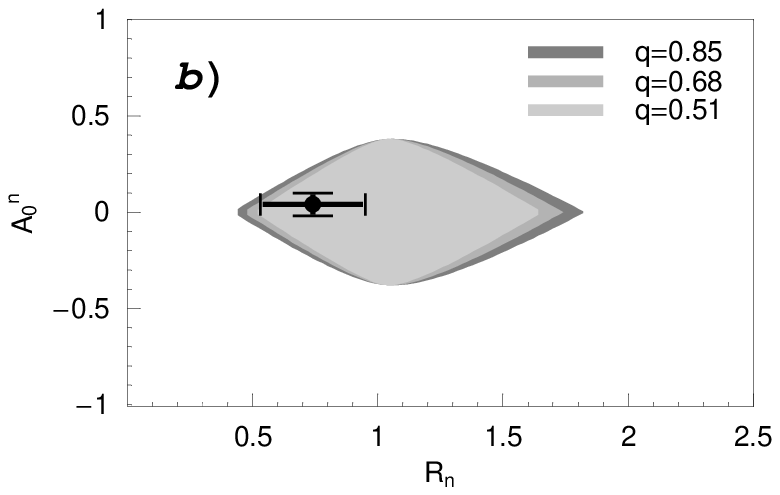}
$$
\vspace*{-0.2cm}
$$\hspace*{-1.cm}
\epsfysize=0.2\textheight
\epsfxsize=0.3\textheight
\epsffile{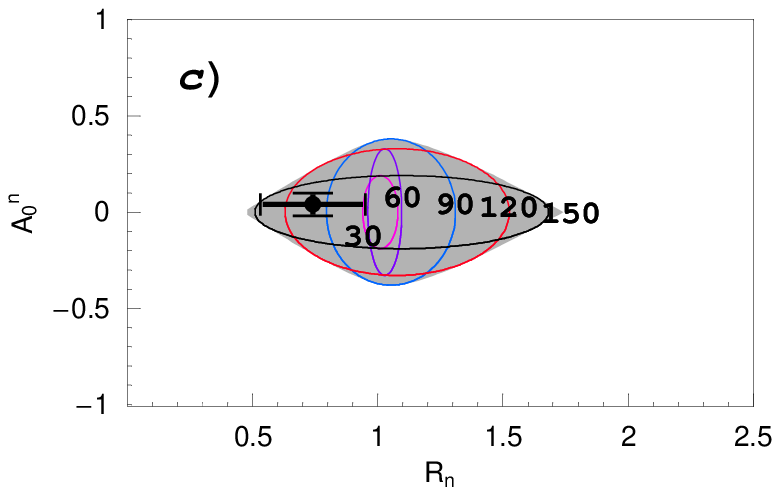} \hspace*{0.3cm}
\epsfysize=0.2\textheight
\epsfxsize=0.3\textheight
\epsffile{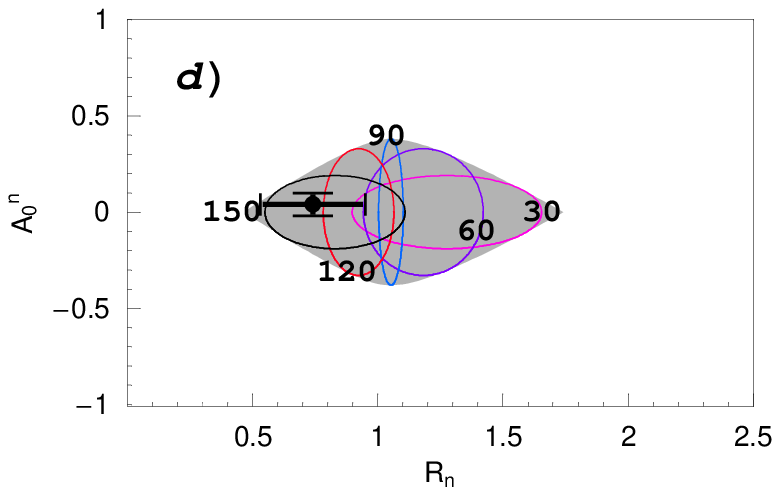}
$$
\vspace*{-0.9cm}
\caption[]{Allowed regions in the $R_{\rm n}$--$A_0^{\rm n}$ plane: 
(a) corresponds to $0.14 \leq r_{\rm n} 
\leq 0.24$ for $q=0.68$, and (b) to $0.51 \leq q \leq 0.85$ for 
$r_{\rm n}=0.19$; the elliptical regions arise if we restrict 
$\gamma$ to the Standard-Model range (\ref{gamma-SM}). In (c) and (d), 
we show the contours for fixed values of $\gamma$ and 
$|\delta_{\rm n}|$, 
respectively ($r_{\rm n}=0.19$, $q=0.68$).}\label{fig:BpiK-neutral}
\end{figure}

\subsection{Numerical Analysis}\label{BpiK-num}
In Figs.~\ref{fig:BpiK-charged} and \ref{fig:BpiK-neutral}, we show the 
allowed regions in observable space of the charged and neutral $B\to\pi K$
systems, respectively. The crosses correspond to the averages of the 
experimental results given in Tables~\ref{tab:BPIK-obs} and 
\ref{tab:BPIK-obs-CPV}, and the elliptical regions arise, if we restrict
$\gamma$ to the Standard-Model range specified in (\ref{gamma-SM}). The 
labels of the contours in (c) refer to the values of $\gamma$ for 
$-180^\circ\leq \delta_{\rm c,n}\leq +180^\circ$, and those of (d) to the 
values of $|\delta_{\rm c,n}|$ for $0^\circ\leq\gamma\leq180^\circ$. Looking 
at these figures, we observe that the experimental data fall pretty well into 
the regions, which are implied by the Standard-Model expressions (\ref{R-expr})
and (\ref{A0-expr}). However, the data points do not favour the restricted 
region, which arises if we constrain $\gamma$ to its Standard-Model range 
(\ref{gamma-SM}). To be more specific, let us consider the contours shown 
in (c) and (d), allowing us to read off the preferred values for $\gamma$ 
and $|\delta_{\rm c,n}|$ directly from the measured observables. In the 
charged $B\to\pi K$ system, the $B$-factory data point towards values for 
$\gamma$ larger than $90^\circ$, and $|\delta_{\rm c}|$ smaller than 
$90^\circ$. In the case of the neutral $B\to\pi K$ system, the data are also 
in favour of $\gamma>90^\circ$, but prefer $|\delta_{\rm n}|$ to be larger 
than $90^\circ$. These features were also pointed out in \cite{BF-neutral2};
in Figs.~\ref{fig:BpiK-charged} and \ref{fig:BpiK-neutral}, we can see 
them directly from the data points. If future measurements should stabilize 
at such a picture, we would have a very exciting situation, since
values for $\gamma$ larger than $90^\circ$ would be in conflict with
the Standard-Model range (\ref{gamma-SM}), and the strong phases
$\delta_{\rm c}$ and $\delta_{\rm n}$ are expected to be of the same 
order of magnitude; factorization would correspond to values around 
$0^\circ$. A possible explanation for such discrepancies would be given 
by large new-physics contributions to the electroweak penguin sector 
\cite{BF-neutral2}. However, it should be kept in mind that we may also
have ``anomalously'' large flavour-symmetry breaking effects. A detailed 
recent analysis of the allowed regions in parameter space of $\gamma$ and 
$\delta_{\rm c,n}$ that are implied by the present $B\to\pi K$ data can 
be found in \cite{ital-corr}, where also very restricted ranges for 
$R_{\rm c,n}$ were obtained by contraining $\gamma$ to its Standard-Model 
expectation. Another $B\to\pi K$ study was recently performed in 
\cite{GR-BpiK-recent}, where the $R_{\rm (c)}$ were calculated for given 
values of $A_0^{\rm (c)}$ as functions of $\gamma$, and were compared with 
the present $B$-factory data.

\begin{figure}
\vspace*{-0.8cm}
$$\hspace*{-1.cm}
\epsfysize=0.2\textheight
\epsfxsize=0.3\textheight
\epsffile{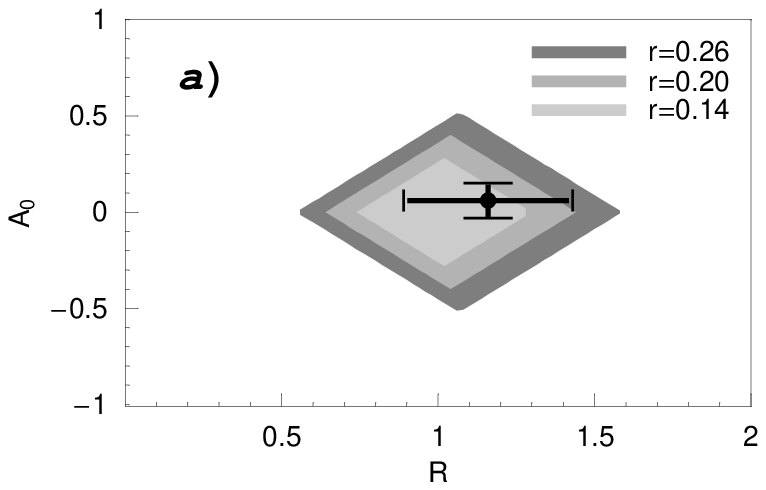} \hspace*{0.3cm}
\epsfysize=0.2\textheight
\epsfxsize=0.3\textheight
\epsffile{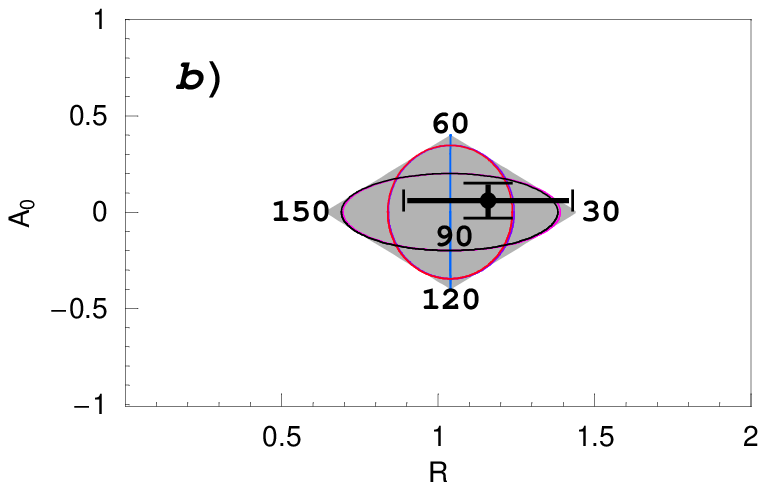}
$$
\vspace*{-0.9cm}
\caption[]{Allowed regions in the $R$--$A_0$ plane: (a) corresponds 
to $0.14 \leq r \leq 0.26$ for $q=0$. In (b), we have chosen $r=0.20$
to show the contours for fixed values of $\gamma$ and $\delta$, which
are identical for $q=0$. Moreover, we obtain the same contours
for $\gamma\to 180^\circ-\gamma$.}\label{fig:BpiK-mixed}
\end{figure}

In Fig.~\ref{fig:BpiK-mixed}, we show the allowed region in observable
space of the mixed $B\to\pi K$ system. Here the crosses represent again 
the averages of the experimental $B$-factory results. Since the expressions 
for $R$ and $A_0$ are symmetric with respect to an interchange of $\gamma$ 
and $\delta$ for $q=0$, the contours for fixed values of $\gamma$ and 
$\delta$ are identical in this limit. Moreover, we obtain the same contours
for $\gamma\to 180^\circ-\gamma$. The experimental data fall well into the 
allowed region, but do not yet allow us to draw any further conclusions. 
In the charged and neutral $B\to\pi K$ systems, the situation appears to 
be much more exciting. 

Let us now turn to the main aspect of our analysis, the $B_d\to\pi^+\pi^-$, 
$B_s\to K^+K^-$ system. In our original paper \cite{Fl-Ma}, we have
addressed these modes only briefly, giving in particular a three-dimensional 
allowed region in the space of the CP asymmetries
${\cal A}_{\rm CP}^{\rm dir}(B_s\to K^+K^-)$, 
${\cal A}_{\rm CP}^{\rm mix}(B_s\to K^+K^-)$ and
${\cal A}_{\rm CP}^{\rm dir}(B_d\to \pi^+\pi^-)$. Here we follow 
\cite{RF-pen-constr}, and use the CP-averaged $B_d\to\pi^\mp K^\pm$ 
branching ratio as an additional input to explore separately the allowed 
regions in the space of the CP-violating $B_d\to\pi^+\pi^-$ and 
$B_s\to K^+K^-$ observables, as well as contraints on $\gamma$. The 
experimental situation has improved significantly since \cite{Fl-Ma} and 
\cite{RF-pen-constr} were written, pointing now to an interesting picture,
although the uncertainties are still too large to draw definite conclusions. 
However, these uncertainties will be reduced considerably in the future due 
to the continuing efforts at the $B$ factories. Once the $B_s\to K^+K^-$
mode is accessible at hadronic $B$ experiments, more refined studies will 
be possible. In the LHC era, the physics potential of the $B_d\to\pi^+\pi^-$, 
$B_s\to K^+K^-$ system can then be fully exploited. In this paper, we
point out that the Standard-Model range in $B_s\to K^+K^-$ observable
space is very constrainted, thereby providing a narrow target range for
these experiments.

\boldmath
\section{Basic Features of the $B_d\to\pi^+\pi^-$, $B_s\to K^+K^-$
\mbox{System} and the Connection with
$B_d\to\pi^\mp K^\pm$}\label{sec:Bdpipi-BsKK}
\unboldmath
\subsection{Amplitude Parametrizations and Observables}
The decay $B_d^0\to\pi^+\pi^-$ originates from $\overline{b}\to\overline{d}$
quark-level transitions. Within the Standard Model, it can be parametrized
as follows \cite{RF-rev}:
\begin{equation}\label{Bd-ampl1}
A(B_d^0\to\pi^+\pi^-)=\lambda_u^{(d)}\left(A_{\rm CC}^{u}+A_{\rm pen}^{u}
\right)+\lambda_c^{(d)}A_{\rm pen}^{c}+\lambda_t^{(d)}A_{\rm pen}^{t}\,,
\end{equation}
where $A_{\rm CC}^{u}$ is due to ``current--current'' contributions, the
amplitudes $A_{\rm pen}^{j}$ describe ``penguin'' topologies with internal
$j$ quarks ($j\in\{u,c,t\})$, and the
\begin{equation}
\lambda_j^{(d)}\equiv V_{jd}V_{jb}^\ast
\end{equation}
are the usual CKM factors. Employing the unitarity of the CKM matrix
and the Wolfenstein parametrization \cite{wolf}, generalized to
include non-leading terms in $\lambda\equiv|V_{us}|=0.222$ \cite{blo},
we arrive at \cite{RF-BsKK}
\begin{equation}\label{Bd-ampl2}
A(B_d^0\to\pi^+\pi^-)={\cal C}\left(e^{i\gamma}-d e^{i\theta}\right),
\end{equation}
where
\begin{equation}\label{Aap-def}
{\cal C}\equiv\lambda^3A R_b\left(A_{\rm CC}^{u}+A_{\rm pen}^{ut}\right),
\end{equation}
with $A_{\rm pen}^{ut}\equiv A_{\rm pen}^{u}-A_{\rm pen}^{t}$, and
\begin{equation}\label{ap-def}
d e^{i\theta}\equiv\frac{1}{R_b}\left(
\frac{A_{\rm pen}^{ct}}{A_{\rm CC}^{u}+A_{\rm pen}^{ut}}\right).
\end{equation}
The quantity $A_{\rm pen}^{ct}$ is defined in analogy to $A_{\rm pen}^{ut}$,
$A\equiv|V_{cb}|/\lambda^2=0.832\pm0.033$, and $R_b$ was already introduced
in (\ref{Rb-def}). The ``penguin parameter'' $d e^{i\theta}$ measures --
sloppily speaking -- the ratio of the $B_d\to\pi^+\pi^-$ ``penguin'' to
``tree'' contributions.

Using the Standard-Model parametrization (\ref{Bd-ampl2}), we obtain
\cite{RF-BsKK}
\begin{eqnarray}
{\cal A}_{\rm CP}^{\rm dir}(B_d\to\pi^+\pi^-)&=&
-\left[\frac{2 d\sin\theta\sin\gamma}{1-
2 d\cos\theta\cos\gamma+d^2}\right]\label{ACP-dir}\\
{\cal A}_{\rm CP}^{\rm mix}(B_d\to\pi^+\pi^-)&=&
\frac{\sin(\phi_d+2\gamma)-2 d \cos\theta \sin(\phi_d+\gamma)+
d^2\sin\phi_d}{1-2 d \cos\theta\cos\gamma+d^2},\label{ACP-mix}
\end{eqnarray}
where $\phi_d=2\beta$ can be determined with the help of (\ref{ACP-BpsiK}),
yielding the twofold solution given in (\ref{phid-det}). Strictly speaking,
mixing-induced CP violation in $B_d\to J/\psi K_{\rm S}$ probes
$\phi_d+\phi_K$, where $\phi_K$ is related to the weak
$K^0$--$\overline{K^0}$ mixing phase and is negligibly small in the Standard
Model. However, due to the small value of the CP-violating parameter
$\varepsilon_K$ of the neutral kaon system, $\phi_K$ can only be affected
by very contrived models of new physics \cite{nirsil}.

In the case of $B_s\to K^+K^-$, we have \cite{RF-BsKK}
\begin{equation}\label{Bs-ampl}
A(B_s^0\to K^+K^-)=\left(\frac{\lambda}{1-\lambda^2/2}\right)
{\cal C}'\left[e^{i\gamma}+\left(\frac{1-\lambda^2}{\lambda^2}\right)
d'e^{i\theta'}\right],
\end{equation}
where
\begin{equation}
{\cal C}'\equiv\lambda^3A\,R_b\left(A_{\rm CC}^{u'}+A_{\rm pen}^{ut'}\right)
\end{equation}
and
\begin{equation}\label{dp-def}
d'e^{i\theta'}\equiv\frac{1}{R_b}
\left(\frac{A_{\rm pen}^{ct'}}{A_{\rm CC}^{u'}+A_{\rm pen}^{ut'}}\right)
\end{equation}
correspond to (\ref{Aap-def}) and (\ref{ap-def}), respectively. The primes
remind us that we are dealing with a $\overline{b}\to\overline{s}$
transition. Introducing
\begin{equation}
\tilde d'\equiv \frac{d'}{\epsilon} \quad \mbox{with} \quad
\epsilon\equiv\frac{\lambda^2}{1-\lambda^2},
\end{equation}
we obtain \cite{RF-BsKK}
\begin{eqnarray}
{\cal A}_{\rm CP}^{\rm dir}(B_s\to K^+K^-)&=&
\frac{2\tilde d'\sin\theta'\sin\gamma}{1+
2\tilde d'\cos\theta'\cos\gamma+\tilde d'^2}\label{ACPs-dir}\\
{\cal A}_{\rm CP}^{\rm mix}(B_s\to K^+K^-)&=&
\frac{\sin(\phi_s+2\gamma)+2\tilde d'\cos\theta'\sin(\phi_s+\gamma)+
\tilde d'^2\sin\phi_s}{1+2\tilde d'\cos\theta'\cos\gamma+
\tilde d'^2},\label{ACPs-mix}
\end{eqnarray}
where the $B^0_s$--$\overline{B^0_s}$ mixing phase
\begin{equation}
\phi_s=-2\lambda^2\eta
\end{equation}
is negligibly small in the Standard Model. Using the range for the
Wolfenstein parameter $\eta$ following from the fits of the unitarity
triangle \cite{UT-fits} yields $\phi_s={\cal O}(-2^\circ)$.
Experimentally, this phase can be probed nicely through $B_s\to J/\psi \phi$,
which allows an extraction of $\phi_s$ also if this phase should be sizeable
due to new-physics contributions to $B^0_s$--$\overline{B^0_s}$ mixing
\cite{nirsil}--\cite{dfn}.

It should be emphasized that (\ref{ACP-dir}), (\ref{ACP-mix}) and
(\ref{ACPs-dir}), (\ref{ACPs-mix}) are completely general parametrizations
of the CP-violating $B_d\to\pi^+\pi^-$ and $B_s\to K^+K^-$ observables,
respectively, relying only on the unitarity of the CKM matrix. If we
assume that $\phi_s$ is negligibly small, as in the Standard Model,
these four observables depend on the four hadronic parameters $d$,
$\theta$, $d'$ and $\theta'$, as well as on the two weak phases $\gamma$
and $\phi_d$. Consequently, we have not sufficient information to
determine these quantities. However, since $B_d\to\pi^+\pi^-$ is related
to $B_s\to K^+K^-$ through an interchange of all down and strange quarks,
the $U$-spin flavour symmetry of strong interactions implies
\begin{equation}\label{U-spin-rel}
d e^{i\theta}=d' e^{i\theta'}.
\end{equation}
Making use of this relation, the parameters $d$, $\theta$, $\gamma$ and
$\phi_d$ can be determined from the CP-violating $B_d\to\pi^+\pi^-$,
$B_s\to K^+K^-$ observables \cite{RF-BsKK}. If we fix $\phi_d$ through
(\ref{ACP-BpsiK}), the use of the $U$-spin symmetry in the extraction
of $\gamma$ can be minimized. Since $d e^{i\theta}$ and $d' e^{i\theta'}$
are defined through ratios of strong amplitudes, the $U$-spin relation
(\ref{U-spin-rel}) is not affected by $U$-spin-breaking corrections in
the factorization approximation \cite{RF-BsKK}, which gives us confidence
in using this relation.
\newpage

\subsection{Constraints on Penguin Parameters}
In order to constrain the hadronic penguin parameters through the
CP-averaged $B_d\to\pi^+\pi^-$ and $B_s\to K^+K^-$ branching ratios, it is
useful to introduce the following quantity \cite{RF-pen-constr}:
\begin{equation}\label{H-def}
H\equiv\frac{1}{\epsilon}\,\left|\frac{{\cal C}'}{{\cal C}}\right|^2
\left[\frac{M_{B_d}}{M_{B_s}}\,\frac{\Phi(M_K/M_{B_s},M_K/M_{B_s})}{
\Phi(M_\pi/M_{B_d},M_\pi/M_{B_d})}\,\frac{\tau_{B_s}}{\tau_{B_d}}\right]
\left[\frac{\mbox{BR}(B_d\to\pi^+\pi^-)}{\mbox{BR}(B_s\to K^+K^-)}\right],
\end{equation}
where
\begin{equation}
\Phi(x,y)\equiv\sqrt{\left[1-(x+y)^2\right]\left[1-(x-y)^2\right]}
\end{equation}
denotes the usual two-body phase-space function. The branching ratio
BR$(B_s\to K^+K^-)$ can be extracted from the ``untagged'' $B_s\to K^+K^-$
rate \cite{RF-BsKK}, where no rapid oscillatory $\Delta M_st$ terms are
present \cite{dunietz-prd}. In the strict $U$-spin limit, we have
\begin{equation}\label{Cp-eq-C}
|{\cal C}'|=|{\cal C}|. 
\end{equation}
Corrections to this relation can be calculated using ``factorization'', 
which yields
\begin{equation}\label{U-fact}
\left|\frac{{\cal C}'}{{\cal C}}\right|_{\rm fact}=\,
\frac{f_K}{f_\pi}\frac{F_{B_sK}(M_K^2;0^+)}{F_{B_d\pi}(M_\pi^2;0^+)}
\left(\frac{M_{B_s}^2-M_K^2}{M_{B_d}^2-M_\pi^2}\right),
\end{equation}
where the form factors $F_{B_sK}(M_K^2;0^+)$ and $F_{B_d\pi}(M_\pi^2;0^+)$
parametrize the hadronic quark-current matrix elements
$\langle K^-|(\bar b u)_{\rm V-A}|B^0_s\rangle$ and
$\langle\pi^-|(\bar b u)_{\rm V-A}|B^0_d\rangle$, respectively \cite{BSW}.
Employing (\ref{Bd-ampl2}) and (\ref{Bs-ampl}) gives
\begin{equation}\label{H-expr}
H=\frac{1-2 d \cos\theta\cos\gamma+d^2}{\epsilon^2+
2 \epsilon d'\cos\theta'\cos\gamma+d'^2}.
\end{equation}
Let us note that there is also an interesting relation between $H$ and the
corresponding direct CP asymmetries \cite{RF-BsKK}:
\begin{equation}\label{CP-rel}
H=-\left(\frac{d\sin\theta}{d'\sin\theta'}\right)\frac{1}{\epsilon}
\left[\frac{{\cal A}_{\rm CP}^{\rm dir}(B_s\to
K^+K^-)}{{\cal A}_{\rm CP}^{\rm dir}(B_d\to\pi^+\pi^-)}\right].
\end{equation}
Relations of this kind are a general feature of $U$-spin-related $B$
decays \cite{gronau-U-spin}.

As can be seen in (\ref{H-expr}), if we use the $U$-spin relation
(\ref{U-spin-rel}), $H$ allows us to determine
\begin{equation}\label{C-def}
C\equiv\cos\theta \cos\gamma
\end{equation}
as a function of $d$ \cite{RF-pen-constr}:
\begin{equation}\label{C-expr}
C=\frac{a-d^2}{2 b d},
\end{equation}
where
\begin{equation}\label{a-b-def}
a\equiv\frac{1-\epsilon^2H}{H-1}\qquad\mbox{and}\qquad
b\equiv\frac{1+\epsilon H}{H-1}.
\end{equation}
Since $C$ is the product of two cosines, it has to satisfy the relation
$-1\leq C\leq +1$, implying the following allowed range for $d$:
\begin{equation}\label{d-bounds1}
\frac{1-\epsilon\sqrt{H}}{1+\sqrt{H}}\leq d\leq\frac{1+\epsilon\sqrt{H}}{|1-
\sqrt{H}|}.
\end{equation}
An alternative derivation of this range, which holds for
$H<1/\epsilon^2=372$, was given in \cite{pirjol}.

\boldmath
\subsection{Connection with $B_d\to\pi^\mp K^\pm$}\label{subsection:conn}
\unboldmath
As we have already noted, experimental data on $B_s\to K^+K^-$ are not
yet available. However, since $B_s\to K^+K^-$ and $B_d\to\pi^\mp K^\pm$
differ only in their spectator quarks, we have
\begin{equation}\label{ACP-rep}
{\cal A}_{\rm CP}^{\rm dir}(B_s\to K^+K^-)\approx{\cal A}_{\rm CP}^{\rm dir}
(B_d\to\pi^\mp K^\pm)
\end{equation}
\begin{equation}\label{BR-rep}
\mbox{BR}(B_s\to K^+K^-)
\approx\mbox{BR}(B_d\to\pi^\mp K^\pm)\,\frac{\tau_{B_s}}{\tau_{B_d}},
\end{equation}
and obtain
\begin{equation}\label{H-res}
H\approx\frac{1}{\epsilon}\left(\frac{f_K}{f_\pi}\right)^2
\left[\frac{\mbox{BR}(B_d\to\pi^+\pi^-)}{\mbox{BR}(B_d\to\pi^\mp K^\pm)}
\right]=\left\{\begin{array}{ll}
7.3\pm2.9 & \mbox{(CLEO \cite{CLEO-BpiK})}\\
9.0\pm1.5 & \mbox{(BaBar \cite{BaBar-Bpipi-new})}\\
8.5\pm3.7 & \mbox{(Belle \cite{belle-BpiK}),}
\end{array}\right.
\end{equation}
yielding the average 
\begin{equation}\label{H-average}
H=8.3\pm1.6,
\end{equation}
which has been calculated by simply adding the errors in quadrature. 
Clearly, the advantage of (\ref{H-res}) is that it allows us to determine
$H$ from the $B$-factory data, without a measurement of $B_s\to K^+K^-$.
On the other hand -- in contrast to (\ref{H-def}) -- this relation relies
not only on $SU(3)$ flavour-symmetry arguments, but also on a certain
dynamical assumption. The point is that $B_s\to K^+K^-$ receives also
contributions from ``exchange'' and ``penguin annihilation'' topologies,
which are absent in $B_d\to\pi^\mp K^\pm$. It is usually assumed that
these contributions play a minor r\^ole \cite{ghlr}. However, they may be
enhanced through certain rescattering effects \cite{res}. The importance of 
the ``exchange'' and ``penguin annihilation'' topologies contributing to
$B_s\to K^+K^-$ can be probed -- in addition to (\ref{ACP-rep}) and
(\ref{BR-rep}) -- with the help of $B_s\to\pi^+\pi^-$. The na\"\i ve 
expectation for the corresponding branching ratio is ${\cal O}(10^{-8})$; 
a significant enhancement would signal that the ``exchange'' and ``penguin 
annihilation'' topologies cannot be neglected. At run II of the Tevatron, 
a first measurement of $B_s\to K^+K^-$ will be possible.

\begin{figure}
\centerline{{
\vspace*{-0.5truecm}
\epsfysize=5.8truecm
{\epsffile{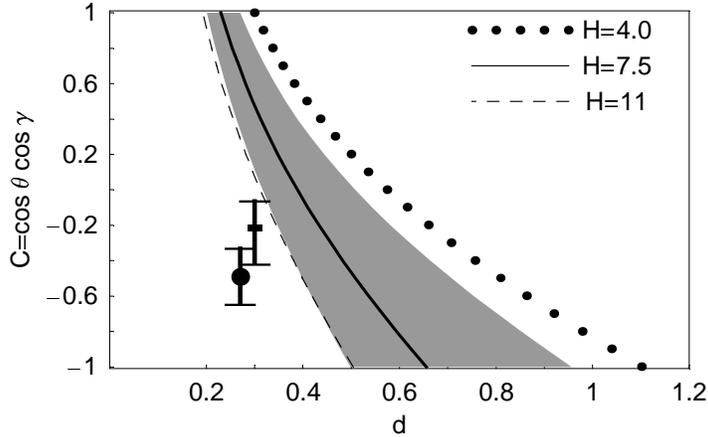}}}}
\caption{The dependence of $C=\cos\theta\cos\gamma$ on $d$ for values of $H$
lying within (\ref{H-range}). The ``circle'' and ``square'' with error 
bars represent the predictions of the QCD factorization \cite{QCD-fact2} and 
PQCD \cite{PQCD-pred} approaches, respectively, for the Standard-Model range 
of $\gamma$ given in (\ref{gamma-SM}). The shaded region corresponds
to a variation of $\xi$ within $[0.8,1.2]$ for $H=7.5$.}\label{fig:d}
\end{figure}

In Fig.~\ref{fig:d}, which is an update of a plot given in
\cite{RF-pen-constr}, we show the dependence of $C$ on $d$ arising
from (\ref{C-expr}) for various values of $H$. Because of possible
uncertainties arising from non-factorizable corrections to 
(\ref{U-fact}) and the dynamical assumptions employed in (\ref{BR-rep}), 
we consider the range
\begin{equation}\label{H-range}
H=7.5\pm3.5,
\end{equation}
which is more conservative than (\ref{H-average}). The 
``circle'' and ``square'' in 
Fig.~\ref{fig:d} represent the predictions for $d e^{i\theta}$ presented
in \cite{QCD-fact2} and \cite{PQCD-pred}, which were obtained within the
QCD factorization \cite{QCD-fact1} and perturbative hard-scattering
(or ``PQCD'') \cite{PQCD} approaches, respectively. The ``error bars''
correspond to the Standard-Model range (\ref{gamma-SM}) for $\gamma$,
whereas the circle and square are evaluated for $\gamma=60^\circ$.
The shaded region in Fig.~\ref{fig:d} corresponds to a variation of
\begin{equation}\label{xi-def}
\xi\equiv d'/d
\end{equation}
within $[0.8,1.2]$ for $H=7.5$. As noted in \cite{RF-pen-constr},
the impact of a sizeable phase difference
\begin{equation}\label{Dtheta-def}
\Delta\theta\equiv\theta'-\theta,
\end{equation}
representing the second kind of possible corrections to (\ref{U-spin-rel}),
is very small in this case.

Looking at Fig.~\ref{fig:d}, we observe that the experimental values for
$H$ imply a rather restricted range for $d$, satisfying $0.2\lsim d\lsim 1$.
Moreover, the curves are not in favour of an interpretation of the QCD
factorization and PQCD predictions for $d e^{i\theta}$ within the Standard
Model. In the latter case, the prediction is somewhat closer to the
``experimental'' curves. This feature is due to the fact that the
CP-conserving strong phase $\theta$ may deviate significantly from its
trivial value of $180^\circ$ in PQCD,
$\theta_{\rm PQCD}=101^\circ\sim 130^\circ$, which is in contrast to the
result of QCD factorization, yielding
$\theta_{\rm QCDF}=185^\circ\sim193^\circ$. As a result, the PQCD approach
may accommodate large direct CP violation in $B_d\to\pi^+\pi^-$, up to the
$50\%$ level \cite{PQCD-pred}, whereas QCD factorization prefers
smaller asymmetries, i.e.\ below the $20\%$ level \cite{QCD-fact2}. In a
recent paper \cite{Neu-Pec}, it was noted that higher-order corrections to
QCD factorization in $B\to\pi K,\pi\pi$ decays may enhance the corresponding
predictions for the CP-conserving strong phases, thereby also
enhancing the direct CP asymmetries. Let us also note that the authors of
\cite{ital-charm-pen}, investigating the impact of ``charming'' penguins
on the QCD factorization approach for $B\to\pi K,\pi\pi$ modes, found values
for ${\cal A}_{\rm CP}^{\rm dir}(B_d\to\pi^+\pi^-)$ as large as
${\cal O}(50\%)$. 

Interestingly, the Belle measurement given in (\ref{Adir-exp}) is actually 
in favour of large direct CP violation in $B_d\to\pi^+\pi^-$. Since we 
restrict $\gamma$ to the range $[0^\circ,180^\circ]$ in our analysis, 
the negative sign of ${\cal A}_{\rm CP}^{\rm dir}(B_d\to\pi^+\pi^-)$ implies 
\begin{equation}\label{theta-range}
0^\circ < \theta < 180^\circ,
\end{equation}
as can be seen in (\ref{ACP-dir}). Interestingly, $\theta_{\rm PQCD}$
is consistent with this range, i.e.\ the sign of the prediction for
${\cal A}_{\rm CP}^{\rm dir}(B_d\to\pi^+\pi^-)$ agrees with the one
favoured by Belle, whereas $\theta_{\rm QCDF}$ lies outside, yielding
the opposite sign for the direct CP asymmetry. 

Another interesting observation in Fig.~\ref{fig:d} is that the theoretical
predictions for the hadronic parameter $d e^{i\theta}$ could be brought to
agreement with the experimental curves for values of $\gamma$
{\it larger} than $90^\circ$ \cite{RF-pen-constr}. In this case, the sign 
of $\cos\gamma$ becomes negative, and the circle and square in 
Fig.~\ref{fig:d} move to positive values of $C$. Arguments for 
$\gamma>90^\circ$ using $B\to PP$, 
$PV$ and $VV$ decays were also given in \cite{HY}. Moreover, as we have
seen in Subsection~\ref{BpiK-num}, the charged and neutral $B\to\pi K$
systems may point towards such values for $\gamma$ as well \cite{BF-neutral2}. 

The constraints arising from $H$ have also implications for the
CP-violating observables of the $B_d\to\pi^+\pi^-$, $B_s\to K^+K^-$
($B_d\to\pi^\mp K^\pm$) decays. In \cite{RF-pen-constr}, upper bounds
on the corresponding direct CP asymmetries and an allowed range for
${\cal A}_{\rm CP}^{\rm mix}(B_d\to\pi^+\pi^-)$ were derived as functions
of $\gamma$. Here we use the information provided by $H$ to explore the
allowed regions in the space of the CP-violating $B_d\to\pi^+\pi^-$ and
$B_s\to K^+K^-$ observables, as well as constraints on $\gamma$. For 
other recent analyses of these decays, we refer the reader to 
\cite{GR-Bpipi,GR-BpiK-recent,GR-BsKK}.

\boldmath
\section{Allowed Regions in $B_d\to\pi^+\pi^-$ Observable
Space}\label{sec:Bdpipi}
\unboldmath
\subsection{General Formulae}\label{subsec:formulae}
The starting point of our considerations is the general expression 
(\ref{ACP-mix}) for ${\cal A}_{\rm CP}^{\rm mix}(B_d\to\pi^+\pi^-)$, 
which allows us to eliminate the strong phase $\theta$ in (\ref{ACP-dir}), 
yielding
\begin{equation}\label{Adir-expr}
{\cal A}_{\rm CP}^{\rm dir}(B_d\to\pi^+\pi^-)=\mp\left[
\frac{\sqrt{4d^2-\left(u+vd^2\right)^2}\sin\gamma}{(1-u\cos\gamma)+
(1-v\cos\gamma)d^2}\right],
\end{equation}
where $u$ and $v$ are defined as in \cite{RF-BsKK}:
\begin{equation}\label{u-def}
u\equiv\frac{{\cal A}_{\rm CP}^{\rm mix}(B_d\to\pi^+\pi^-)-
\sin(\phi_d+2\gamma)}{{\cal A}_{\rm CP}^{\rm mix}(B_d\to\pi^+\pi^-)\cos\gamma
-\sin(\phi_d+\gamma)}
\end{equation}
\begin{equation}\label{v-def}
v\equiv\frac{{\cal A}_{\rm CP}^{\rm mix}(B_d\to\pi^+\pi^-)-
\sin\phi_d}{{\cal A}_{\rm CP}^{\rm mix}(B_d\to\pi^+\pi^-)\cos\gamma
-\sin(\phi_d+\gamma)}.
\end{equation}
It should be emphasized that (\ref{Adir-expr}) is valid exactly. If we use the
$U$-spin relation (\ref{U-spin-rel}), we may also eliminate $\theta$ through
${\cal A}_{\rm CP}^{\rm mix}(B_d\to\pi^+\pi^-)$ in (\ref{H-expr}).
Taking into account, moreover, the possible corrections to
(\ref{U-spin-rel}) through (\ref{xi-def}) and (\ref{Dtheta-def}), we
obtain the following expression for $d^2$:
\begin{equation}\label{d2-det}
d^2=\frac{AB+(2-uv)S^2\pm |S|\sqrt{4AB-(Av+Bu)^2+4(1-uv)S^2}}{B^2+v^2S^2},
\end{equation}
where
\begin{eqnarray}
A&\equiv&1-\epsilon^2H-u\left(1+\epsilon\xi H\cos\Delta\theta\right)
\cos\gamma\\
B&\equiv&\xi^2H-1+v\left(1+\epsilon\xi H\cos\Delta\theta\right)\cos\gamma\\
S&\equiv&\epsilon\xi H \cos\gamma\sin\Delta\theta.
\end{eqnarray}
In the limit of $\Delta\theta=0^\circ$, (\ref{d2-det}) simplifies to
\begin{equation}
\left.d^2\right|_{\Delta\theta=0^\circ}=\frac{A}{B}=
\frac{1-\epsilon^2H-u\left(1+\epsilon\xi H\right)\cos\gamma}{\xi^2H-1+
v\left(1+\epsilon\xi H\right)\cos\gamma}.
\end{equation}
If we now insert $d^2$ thus determined into (\ref{Adir-expr}), we may 
calculate ${\cal A}_{\rm CP}^{\rm dir}(B_d\to\pi^+\pi^-)$ as a function
of $\gamma$ for given values of $H$,
${\cal A}_{\rm CP}^{\rm mix}(B_d\to\pi^+\pi^-)$ and $\phi_d$. It is an
easy exercise to show that (\ref{Adir-expr}) and (\ref{d2-det}) are
invariant under the following replacements:
\begin{equation}\label{sym-rel}
\phi_d\to 180^\circ-\phi_d,\quad \gamma\to 180^\circ-\gamma,
\end{equation}
which will have important consequences below. 

In the following, we assume that $\phi_d$ and $H$ are known from
(\ref{phid-det}) and (\ref{H-res}), respectively. If we then
vary $\gamma$ within $[0^\circ,180^\circ]$ for each value of 
${\cal A}_{\rm CP}^{\rm mix}(B_d\to\pi^+\pi^-)\in[-1,+1]$,
we obtain an allowed range in the ${\cal A}_{\rm CP}^{\rm mix}
(B_d\to\pi^+\pi^-)$--${\cal A}_{\rm CP}^{\rm dir}(B_d\to\pi^+\pi^-)$
plane. Restricting $\gamma$ to (\ref{gamma-SM}), a more constrained region
arises. The allowed range in the ${\cal A}_{\rm CP}^{\rm mix}
(B_d\to\pi^+\pi^-)$--${\cal A}_{\rm CP}^{\rm dir}(B_d\to\pi^+\pi^-)$
plane can be obtained alternatively by eliminating
$d$ through $H$ in (\ref{ACP-dir}) and (\ref{ACP-mix}), and then varying
$\gamma$ and $\theta$ within the ranges of $[0^\circ,180^\circ]$ and
$[-180^\circ,+180^\circ]$, respectively. 

A different approach to analyse the situation in the ${\cal A}_{\rm CP}^{\rm 
mix}(B_d\to\pi^+\pi^-)$--${\cal A}_{\rm CP}^{\rm dir}(B_d\to\pi^+\pi^-)$ 
plane was employed in \cite{GR-Bpipi}. In this paper, the parameter 
$de^{i\theta}$ introduced in (\ref{ap-def}) is written as 
$-P_{\pi\pi}/T_{\pi\pi}$, where the magnitude of the ``penguin'' amplitude 
$P_{\pi\pi}$ is fixed through the CP-averaged branching ratio of the 
penguin-dominated decay $B^\pm\to\pi^\pm K$ with the help of $SU(3)$ 
flavour-symmetry arguments and plausible dynamical assumptions, concerning
the neglect of an annihilation amplitude ${\cal A}$. In order to deal with
$T_{\pi\pi}\propto(A^u_{\rm CC}+A^{ut}_{\rm pen})$, the penguin piece
$A^{ut}_{\rm pen}$ is neglected, and the magnitude of the ``tree'' 
amplitude $A^u_{\rm CC}$ is estimated using factorization and data on 
$B\to\pi\ell\nu$, yielding $d\equiv|P_{\pi\pi}/T_{\pi\pi}|=0.276\pm0.064$ 
\cite{GR-BpiK-recent}.\footnote{The dynamical assumptions concerning 
${\cal A}$ and $A^{ut}_{\rm pen}$ may be affected by large
rescattering effects \cite{res}.} Moreover, using the unitarity relation 
$\gamma=180^\circ-\alpha-\beta$ to eliminate $\gamma$, the observables 
${\cal A}_{\rm CP}^{\rm mix}(B_d\to\pi^+\pi^-)$ and 
${\cal A}_{\rm CP}^{\rm dir}(B_d\to\pi^+\pi^-)$ are expressed in terms of 
$\alpha$, $\beta$ and $P_{\pi\pi}/T_{\pi\pi}$. Fixing
$\beta$ to be equal to the ``Standard-Model'' solution of $26^\circ$ implied
by $B_d\to J/\psi K_{\rm S}$, and estimating $|P_{\pi\pi}/T_{\pi\pi}|$ as 
sketched above, ${\cal A}_{\rm CP}^{\rm mix}(B_d\to\pi^+\pi^-)$ and 
${\cal A}_{\rm CP}^{\rm dir}(B_d\to\pi^+\pi^-)$ depend only on $\alpha$
and $\theta$. For each given value of $\alpha$, the variation of $\theta$ 
within the range $[-180^\circ,+180^\circ]$ specifies then a contour in the
${\cal A}_{\rm CP}^{\rm mix}(B_d\to\pi^+\pi^-)$--${\cal A}_{\rm CP}^{\rm 
dir}(B_d\to\pi^+\pi^-)$ plane, holding within the Standard Model. 

In our analysis, we prefer to use $H$ as an additional observable to deal 
with the penguin contributions, i.e.\ with the parameter $de^{i\theta}$,
since we have then not to make a separation between $P_{\pi\pi}$ and 
$T_{\pi\pi}$, and have in particular not to rely on the na\"\i ve 
factorization approach to estimate the overall magnitude of $T_{\pi\pi}$, 
which is governed by colour-allowed tree-diagram-like processes, but may
also be affected by penguin contributions. In our approach,
factorization is only used to include $SU(3)$-breaking effects. Concerning
the parametrization in terms of weak phases, we prefer to use $\gamma$ and 
the general $B^0_d$--$\overline{B^0_d}$ mixing phase $\phi_d$, since the 
results for the former quantity can then be compared easily with 
constraints from other processes, whereas the latter can anyway be fixed 
straighforwardly through mixing-induced CP violation in 
$B_d\to J/\psi K_{\rm S}$ up to a twofold ambiguity, also if there should 
be CP-violating new-physics contributions to $B^0_d$--$\overline{B^0_d}$ 
mixing. This way, we obtain an interesting connection between the two 
solutions for $\phi_d$ and the allowed ranges for $\gamma$, as we will 
see in the next subsection.

\subsection{Numerical Analysis}\label{subsec:numerics}
\begin{figure}
\vspace*{-0.5cm}
$$\hspace*{-1.cm}
\epsfysize=0.2\textheight
\epsfxsize=0.3\textheight
\epsffile{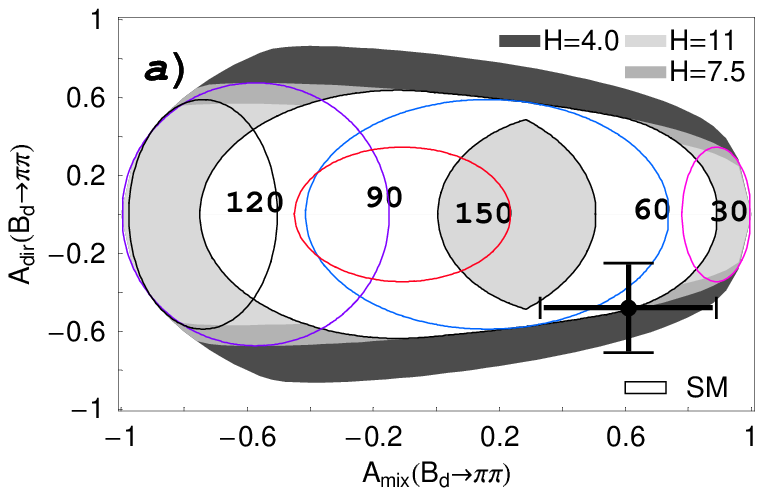} \hspace*{0.3cm}
\epsfysize=0.2\textheight
\epsfxsize=0.3\textheight
\epsffile{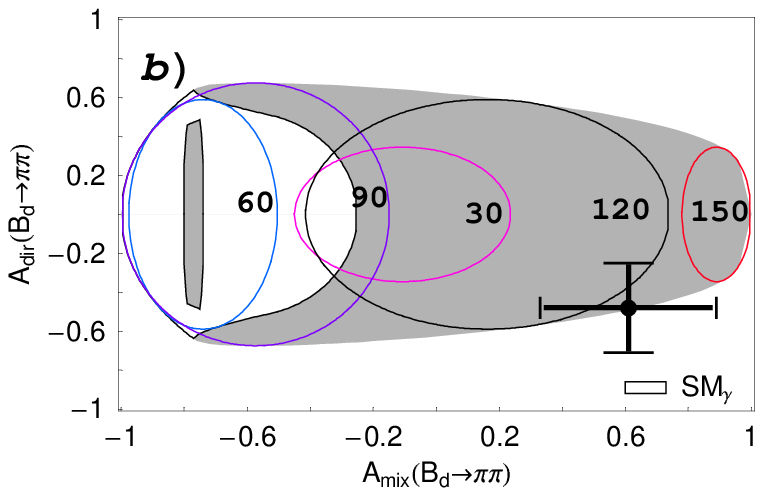}
$$
\vspace*{-0.9cm}
\caption[]{Allowed region in the 
${\cal A}_{\rm CP}^{\rm mix}(B_d\to\pi^+\pi^-)$--${\cal A}_{\rm CP}^{\rm
dir}(B_d\to\pi^+\pi^-)$ plane for (a) $\phi_d=51^\circ$ and various values
of $H$, and (b) $\phi_d=129^\circ$ and $H=7.5$. The SM regions arise if we 
restrict $\gamma$ to (\ref{gamma-SM}) ($H=7.5$). We have also included 
the contours arising for various fixed values of $\gamma$.}\label{fig:AdAmpipi}
\end{figure}

In Fig.~\ref{fig:AdAmpipi}, we show the situation in the
${\cal A}_{\rm CP}^{\rm mix}(B_d\to\pi^+\pi^-)$--${\cal A}_{\rm CP}^{\rm 
dir}(B_d\to\pi^+\pi^-)$ plane for the central values of the two solutions
for $\phi_d$ given in (\ref{phid-det}), and values of $H$ lying within
(\ref{H-range}). The impact of the present experimental uncertainty of 
$\phi_d$ is already very small, and will become negligible in the future. 
In order to calculate Fig.~\ref{fig:AdAmpipi}, we have used, for simplicity, 
$\xi=1$ and $\Delta \theta=0^\circ$; the impact of variations of these 
parameters will be discussed in Subsection~\ref{subsec:xi-Dtheta}. The 
contours in Fig.~\ref{fig:AdAmpipi} arise, if we fix $\gamma$ to the values 
specified through the labels, and vary $\theta$ within 
$[-180^\circ,+180^\circ]$. We have also indicated the region which arises 
if we restrict $\gamma$ to the Standard-Model range (\ref{gamma-SM}). The
crosses describe the experimental averages given in 
(\ref{CP-Bpipi-average}).

We observe that the experimental averages overlap -- within their 
uncertainties -- nicely with the SM region for $\phi_d=51^\circ$, and
point towards $\gamma\sim50^\circ$. In this case, not only $\gamma$ 
would be in accordance with the results of the fits of the unitarity 
triangle \cite{UT-fits}, but also the $B^0_d$--$\overline{B^0_d}$ mixing 
phase $\phi_d$. On the other hand, for $\phi_d=129^\circ$, the experimental 
values favour $\gamma\sim130^\circ$, and have essentially no overlap with 
the SM region. This feature is due to the symmetry relations given in
(\ref{sym-rel}). Since a value of $\phi_d=129^\circ$ would require
CP-violating new-physics contributions to $B^0_d$--$\overline{B^0_d}$ 
mixing, also the $\gamma$ range in (\ref{gamma-SM}) may no longer hold, 
as it relies strongly on a Standard-Model interpretation of the experimental 
information on 
$B^0_{d,s}$--$\overline{B^0_{d,s}}$ mixing 
\cite{UT-fits}. In particular, also values for $\gamma$ larger than 
$90^\circ$ could then in principle be accommodated. As we have noted
in Subsection~\ref{subsection:conn}, theoretical analyses of $d e^{i\theta}$
would actually favour values for $\gamma$ being larger than $90^\circ$, 
provided that the corresponding theoretical uncertainties are reliably under 
control, and that the $B_d\to\pi^+\pi^-$, $B_s\to K^+K^-$ system is still 
described by the Standard-Model parametrizations. In this case, 
(\ref{CP-Bpipi-average}) would point towards a $B^0_d$--$\overline{B^0_d}$ 
mixing phase of $129^\circ$, which would be a very exciting situation. 

Consequently, it is very important to resolve the twofold ambiguity 
arising in (\ref{phid-det}) directly. To this end, $\cos\phi_d$ has to 
be measured as well. For the resolution of the discrete
ambiguity, already a determination of the sign of $\cos\phi_d$ would be 
sufficient, where a positive result would imply that $\phi_d$ is
given by $51^\circ$. There are several strategies on the market to accomplish 
this goal \cite{dfn,ambig}. Unfortunately, they are challenging from an 
experimental point of view and will require a couple of years of taking 
further data at the $B$ factories.

\begin{figure}
\vspace*{-0.5cm}
$$\hspace*{-1.cm}
\epsfysize=0.2\textheight
\epsfxsize=0.3\textheight
\epsffile{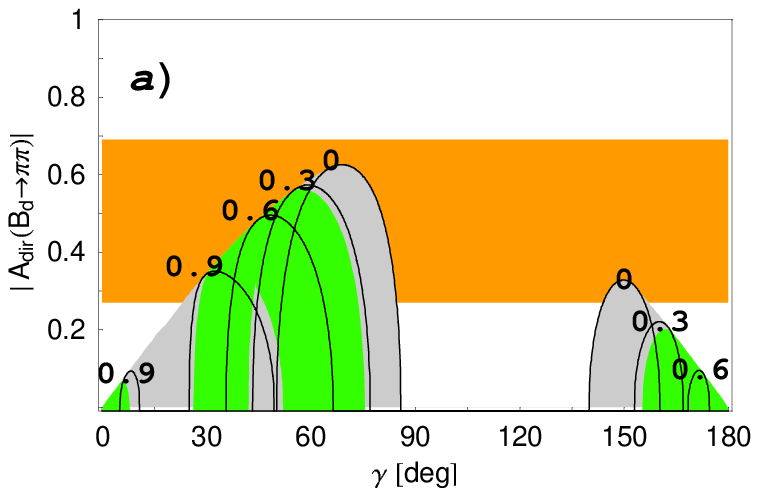} \hspace*{0.3cm}
\epsfysize=0.2\textheight
\epsfxsize=0.3\textheight
\epsffile{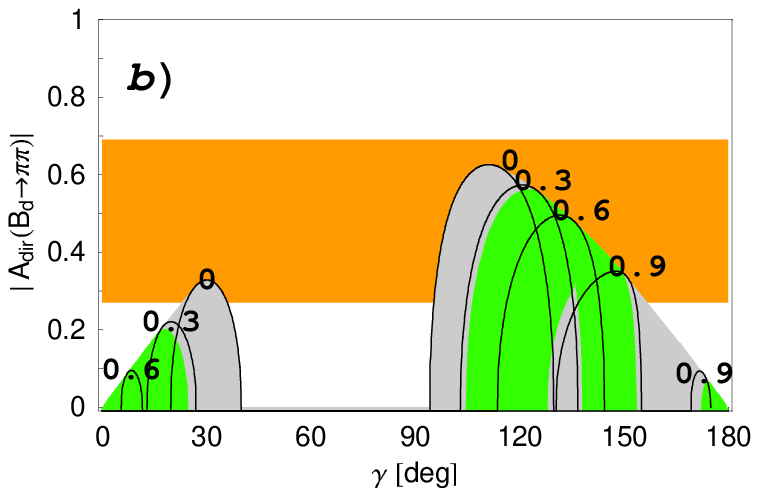}
$$
\vspace*{-0.5cm}
$$\hspace*{-1.cm}
\epsfysize=0.2\textheight
\epsfxsize=0.3\textheight
\epsffile{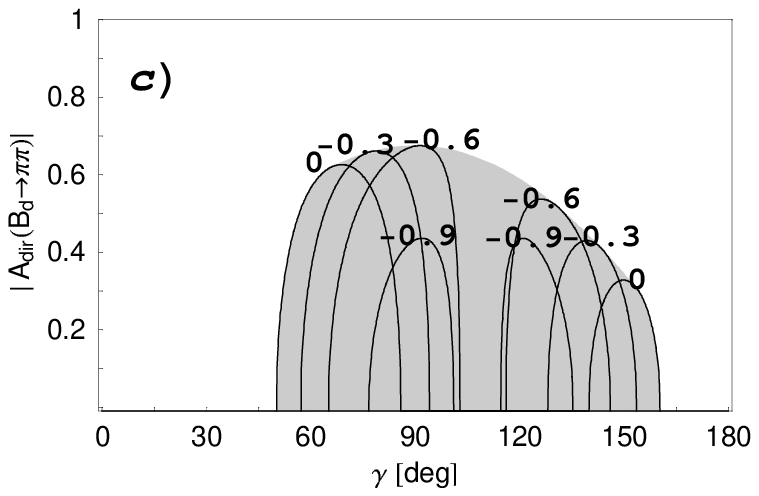} \hspace*{0.3cm}
\epsfysize=0.2\textheight
\epsfxsize=0.3\textheight
 \epsffile{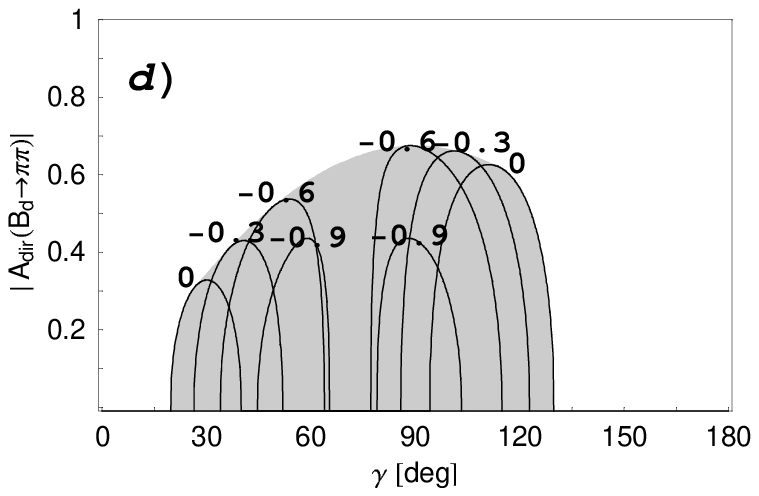}
$$
\vspace*{-0.9cm}
\caption[]{Dependence of $|{\cal A}_{\rm CP}^{\rm dir}(B_d\to\pi^+\pi^-)|$
on $\gamma$ for various values of 
${\cal A}_{\rm CP}^{\rm mix}(B_d\to\pi^+\pi^-)$ in the case of $H=7.5$. 
In (a) and (b), we have chosen $\phi_d=51^\circ$ and $\phi_d=129^\circ$,
respectively. The shaded region arises from a variation of 
${\cal A}_{\rm CP}^{\rm mix}(B_d\to\pi^+\pi^-)$ within $[0,+1]$. The
corresponding plots for negative  
${\cal A}_{\rm CP}^{\rm mix}(B_d\to\pi^+\pi^-)$ are shown in (c) and
(d) for $\phi_d=51^\circ$ and $\phi_d=129^\circ$, respectively. We have also
included the bands arising from the experimental averages in 
(\ref{CP-Bpipi-average}).}\label{fig:gam-Add}
\end{figure}

In order to put these observations on a more quantitative basis, 
we show in Fig.~\ref{fig:gam-Add} 
the dependences of $|{\cal A}_{\rm CP}^{\rm dir}(B_d\to\pi^+\pi^-)|$ on 
$\gamma$ for given values of ${\cal A}_{\rm CP}^{\rm mix}(B_d\to\pi^+\pi^-)$. 
For the two solutions of $\phi_d$, an interesting difference arises, if
we consider positive and negative values of the mixing-induced CP asymmetry, 
as done in (a), (b) and (c), (d), respectively, In the former case,
we obtain the following {\it excluded} ranges for $\gamma$:
\begin{equation}
86^\circ\lsim\gamma\lsim140^\circ \, (\phi_d=51^\circ), \quad
40^\circ\lsim\gamma\lsim94^\circ \, (\phi_d=129^\circ).
\end{equation}
Consequently, for $\phi_d=51^\circ$, we can conveniently accommodate the
Standard-Model range (\ref{gamma-SM}), in contrast to the situation
for $\phi_d=129^\circ$. On the other hand, if we consider negative values 
of ${\cal A}_{\rm CP}^{\rm mix}(B_d\to\pi^+\pi^-)$, we obtain the
following {\it allowed} ranges for $\gamma$:
\begin{equation}
50^\circ\lsim\gamma\lsim160^\circ \, (\phi_d=51^\circ), \quad
20^\circ\lsim\gamma\lsim130^\circ \, (\phi_d=129^\circ).
\end{equation}
In this case, both ranges would contain (\ref{gamma-SM}), and the
situation would not be as exciting as for a positive value of 
${\cal A}_{\rm CP}^{\rm mix}(B_d\to\pi^+\pi^-)$. These features can be
understood in a rather transparent manner from the extremal values for
${\cal A}_{\rm CP}^{\rm mix}(B_d\to\pi^+\pi^-)$ derived in 
\cite{RF-pen-constr}.

In Fig.~\ref{fig:gam-Add}, we have also included bands, which are due 
to the present experimental averages given in (\ref{CP-Bpipi-average}). 
Interestingly, a positive value of 
${\cal A}_{\rm CP}^{\rm mix}(B_d\to\pi^+\pi^-)$ is now favoured by the
data. From the overlap of the 
${\cal A}_{\rm CP}^{\rm mix}(B_d\to\pi^+\pi^-)$ and
$|{\cal A}_{\rm CP}^{\rm dir}(B_d\to\pi^+\pi^-)|$ bands we obtain
the following solutions for $\gamma$:
\begin{equation}\label{gam-res}
28^\circ\lsim\gamma\lsim74^\circ \, (\phi_d=51^\circ), \quad
106^\circ\lsim\gamma\lsim152^\circ \, (\phi_d=129^\circ).
\end{equation}
In the future, the experimental uncertainties will be reduced considerably, 
thereby providing much more stringent results for $\gamma$. Moreover, it
should be emphasized that also $d$ can be determined with the help of 
(\ref{d2-det}). Going then back to (\ref{C-expr}), we may extract 
$\cos\theta$ as well, which allows an unambiguous determination of 
$\theta$ because of (\ref{theta-range}). Before we come back to this issue 
in Subsection~\ref{subsec:fact}, where we shall have a brief look at 
factorization and the discrete ambiguities arising typically in the 
extraction of $\gamma$ from the contours shown in Fig.~\ref{fig:gam-Add}, 
let us first turn to the uncertainties associated with the parameters 
$\xi$ and $\Delta\theta$.

\begin{figure}[t]
\vspace*{-0.2cm}
$$\hspace*{-0.5cm}
\epsfysize=0.2\textheight
\epsfxsize=0.3\textheight
\epsffile{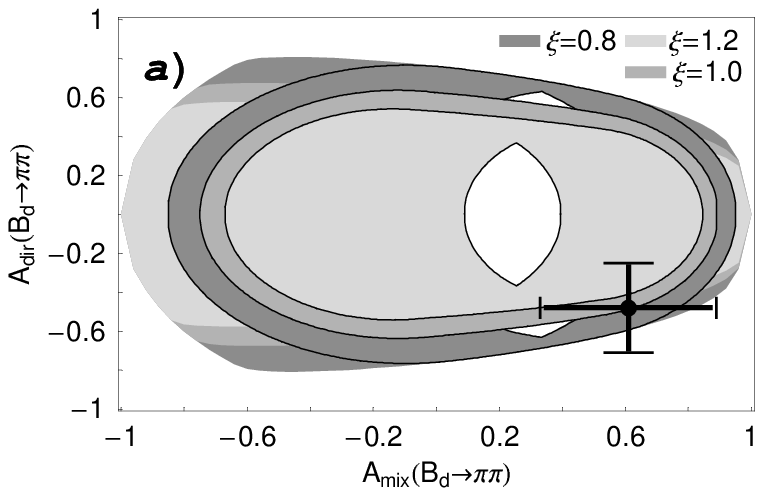} \hspace*{0.3cm}
\epsfysize=0.2\textheight
\epsfxsize=0.3\textheight
\epsffile{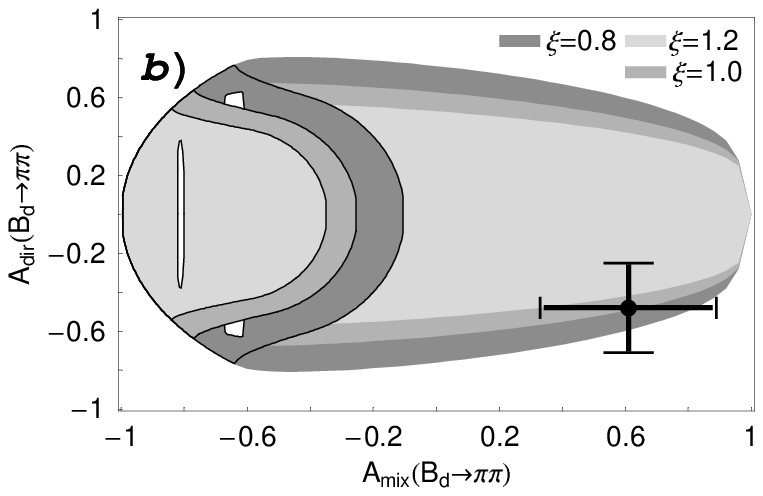}
$$
\vspace*{-0.5cm}
$$\hspace*{-1.cm}
\epsfysize=0.2\textheight
\epsfxsize=0.3\textheight
\epsffile{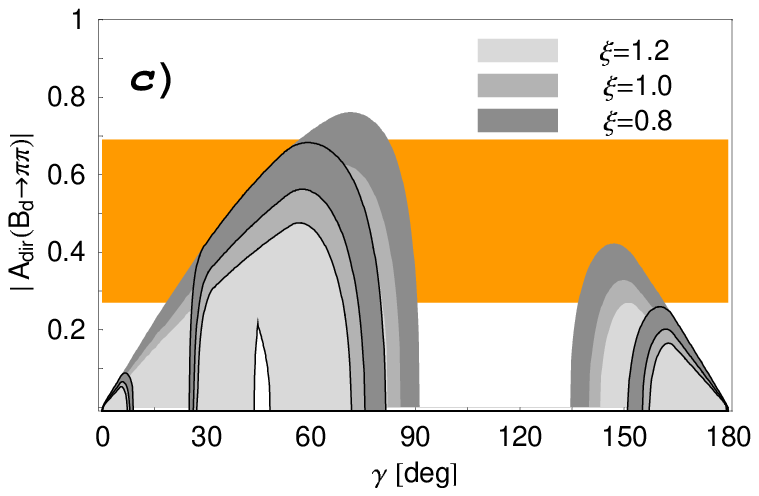} \hspace*{0.3cm}
\epsfysize=0.2\textheight
\epsfxsize=0.3\textheight
\epsffile{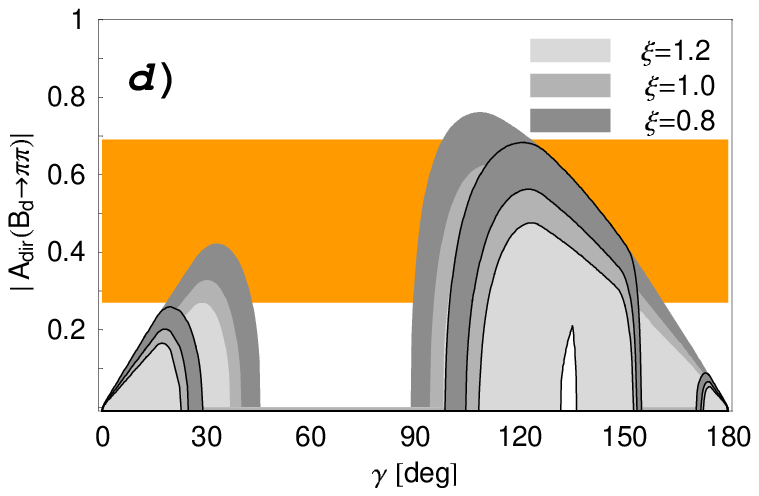}
$$
\vspace*{-0.9cm}
\caption[]{Impact of a variation of $\xi$ within $[0.8,1.2]$ for $H=7.5$
on (a), (b) the allowed ranges in the ${\cal A}_{\rm CP}^{\rm mix}
(B_d\to\pi^+\pi^-)$--${\cal A}_{\rm CP}^{\rm dir}(B_d\to\pi^+\pi^-)$
plane, and (c), (d) the 
$\gamma$--$|{\cal A}_{\rm CP}^{\rm dir}(B_d\to\pi^+\pi^-)|$ plane for
positive values of ${\cal A}_{\rm CP}^{\rm 
mix}(B_d\to\pi^+\pi^-)$, as explained in the text. 
We have used $\phi_d=51^\circ$ and $\phi_d=129^\circ$ in (a), (c) and 
(b), (d), respectively.}\label{fig:Bpipi-xi}
\end{figure}

\newpage

\boldmath
\subsection{Sensitivity on $\xi$ and $\Delta\theta$}\label{subsec:xi-Dtheta}
\unboldmath
In the numerical analysis discussed in Subsection~\ref{subsec:numerics}, we 
have used $\xi=1$ and $\Delta\theta=0^\circ$. Let us now investigate the 
sensitivity of our results on deviations of $\xi$ from 1, and sizeable values 
of $\Delta\theta$. The formulae given in Subsection~\ref{subsec:formulae}
take into account these parameters exactly, thereby allowing us to study
their effects straightforwardly. It turns out that the impact of 
$\Delta\theta$ is very small,\footnote{We shall give a plot illustrating 
the impact of $\Delta\theta\not=0^\circ$ on the $B_s\to K^+K^-$ analysis in 
Subsection~\ref{subsec:BsKK-num}.} even for values as large as $\pm20^\circ$. 
Consequently, the most important effects are due to the parameter $\xi$. 
In Fig.~\ref{fig:Bpipi-xi}, we use $H=7.5$ to illustrate the impact of a 
variation of $\xi$ within the range $[0.8,1.2]$: in (a) and (b), we show 
the allowed region in the 
${\cal A}_{\rm CP}^{\rm mix}(B_d\to\pi^+\pi^-)$--${\cal A}_{\rm CP}^{\rm 
dir}(B_d\to\pi^+\pi^-)$ plane for $\phi_d=51^\circ$ and $129^\circ$, 
respectively, including also the regions, which arise if we restrict
$\gamma$ to the Standard-Model range (\ref{gamma-SM}). In (c) and 
(d), we show the corresponding situation in the 
$\gamma$--$|{\cal A}_{\rm CP}^{\rm dir}(B_d\to\pi^+\pi^-)|$ plane for
positive values of ${\cal A}_{\rm CP}^{\rm mix}(B_d\to\pi^+\pi^-)$. Here
we have also included the bands arising from the present experimental 
values for CP violation in $B_d\to\pi^+\pi^-$. We find that a variation
of $\xi$ within $[0.8,1.2]$ affects our result (\ref{gam-res}) for $\gamma$
as follows:
\begin{equation}\label{gam-res-xi}
(28\pm2)^\circ\lsim\gamma\lsim(74\pm6)^\circ \, (\phi_d=51^\circ), \quad
(106\pm6)^\circ\lsim\gamma\lsim(152\pm2)^\circ \, (\phi_d=129^\circ).
\end{equation}
For future reduced experimental uncertainties of 
$|{\cal A}_{\rm CP}^{\rm dir}(B_d\to\pi^+\pi^-)|$, also the holes in 
Figs.~\ref{fig:Bpipi-xi} (c) and (d) may have an impact on $\gamma$, 
excluding certain values. The impact of the hole is increasing for decreasing 
values of $\xi$. In Figs.~\ref{fig:Bpipi-xi} (c) and (d),
only the smallest holes for $\xi=1.2$ are shown, whereas those corresponding 
$\xi=1.0$ and $\xi=0.8$ are hidden.

The range for $\xi$ considered in Figs.~\ref{fig:d} and \ref{fig:Bpipi-xi}
appears rather conservative to us, since (\ref{U-spin-rel}) is not 
affected by $U$-spin-breaking corrections within 
the factorization approach, in contrast to (\ref{Cp-eq-C}), as can be seen
in (\ref{U-fact}). Non-factorizable corrections to the latter 
relation would show up as a systematic shift of $H$, and could be taken 
into account straightforwardly in our formalism.

\begin{figure}
\vspace*{-0.2cm}
$$\hspace*{-1.cm}
\epsfysize=0.2\textheight
\epsfxsize=0.3\textheight
\epsffile{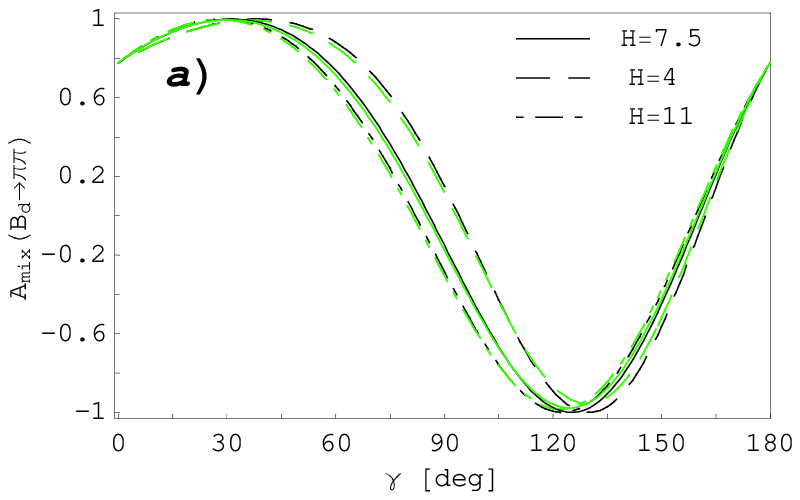} \hspace*{0.3cm}
\epsfysize=0.2\textheight
\epsfxsize=0.3\textheight
 \epsffile{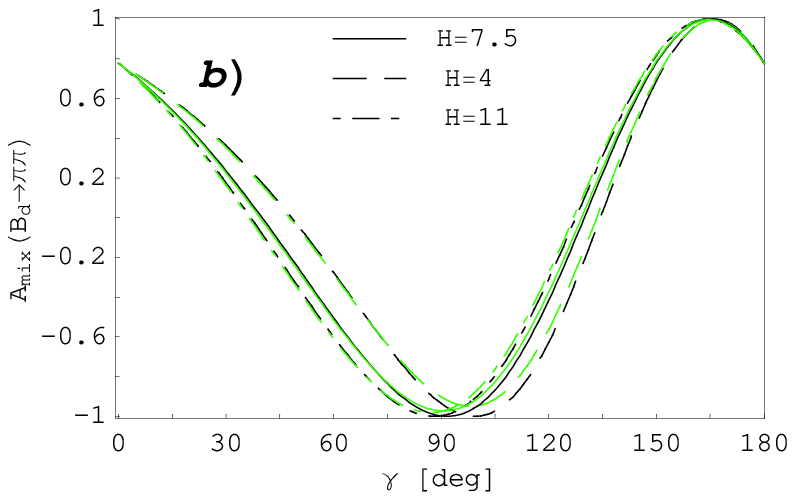}
$$
\vspace*{-0.9cm}
\caption{The dependence of ${\cal A}_{\rm CP}^{\rm mix}(B_d\to\pi^+\pi^-)$
on $\gamma$ arising from (\ref{ACP-mix-fact}) with (\ref{d-expr-fact})
for $c=c'=1$. In (a) and (b), we have chosen $\phi_d=51^\circ$ 
and $\phi_d=129^\circ$, respectively. We have also indicated the 
small shifts of the curves for a variation of $\theta=\theta'$
between $160^\circ$ and $200^\circ$.}\label{fig:fact1}
\end{figure}

\boldmath
\subsection{Comments on Factorization and 
Discrete Ambiguities}\label{subsec:fact}
\unboldmath
As we have noted in the introduction, the present BaBar and Belle 
measurements of CP violation in $B_d\to\pi^+\pi^-$ are not fully
consistent with each other. Whereas Belle is in favour of very large
CP asymmetries in this channel, the central values obtained by 
BaBar are close to zero. The Belle result and the average for 
${\cal A}_{\rm CP}^{\rm dir}(B_d\to\pi^+\pi^-)$ given in 
(\ref{CP-Bpipi-average}) cannot be accommodated within the 
factorization picture, predicting $\theta\sim180^\circ$. On the other
hand, this framework would still be consistent with BaBar. Let us therefore 
spend a few words on simplifications of the analysis given above that can
be obtained by using a rather mild input from factorization. 

If we look at (\ref{ACP-mix}) and (\ref{H-expr}), we observe that 
${\cal A}_{\rm CP}^{\rm mix}(B_d\to\pi^+\pi^-)$ and $H$ depend only 
on cosines of strong phases, which would be equal to $-1$
within factorization. In contrast to $\sin\theta$, the value of 
$\cos\theta$ is not very sensitive to deviations of $\theta$ from 
$\left.\theta\right|_{\rm fact}\sim180^\circ$, i.e.\ to 
non-factorizable effects. Using (\ref{H-expr}), we obtain
\begin{equation}\label{d-expr-fact}
d=\tilde b\cos\gamma+\sqrt{\tilde a+\left(\tilde b\cos\gamma\right)^2},
\end{equation}
where
\begin{equation}
\tilde a\equiv\frac{1-\epsilon^2H}{\xi^2H-1}, \quad
\tilde b\equiv\frac{c+c' \epsilon\xi H}{\xi^2H-1}
\end{equation}
with
\begin{equation}\label{c-cp-def}
c\equiv-\cos\theta, \quad c'\equiv-\cos\theta' 
\end{equation}
are generalizations of $a$ and $b$ introduced in (\ref{a-b-def}).
The parameters $c$ and $c'$ allow us to take into account deviations from
the strict factorization limit, implying $c=c'=1$. We may now calculate
\begin{equation}\label{ACP-mix-fact}
{\cal A}_{\rm CP}^{\rm mix}(B_d\to\pi^+\pi^-)=
\frac{\sin(\phi_d+2\gamma)+2dc\sin(\phi_d+\gamma)+d^2\sin\phi_d}{
1+2dc\cos\gamma+d^2}
\end{equation}
with the help of (\ref{d-expr-fact}) as a function of $\gamma$. 

In Fig.~\ref{fig:fact1}, we show the corresponding curves for various 
values of $H$ in the case of $c=c'=1$; we have again to distinguish 
between (a) $\phi_d=51^\circ$, and (b) $\phi_d=129^\circ$. For a
variation of $\theta=\theta'$ between $160^\circ$ and $200^\circ$,
we obtain very small shifts of these curves, as indicated in the figure.
For ${\cal A}_{\rm CP}^{\rm mix}(B_d\to\pi^+\pi^-)\sim0$, as favoured
by the present BaBar result, we would obtain
\begin{equation}\label{gamma-fact}
\gamma\sim86^\circ \lor 160^\circ \,\, (\phi_d=51^\circ), \quad
\gamma\sim40^\circ \lor 130^\circ \,\, (\phi_d=129^\circ).
\end{equation}
Using now once more (\ref{d-expr-fact}) or the curves shown in 
Fig.~\ref{fig:d} yields 
correspondingly
\begin{equation}\label{D-fact-ranges}
d\sim0.4 \lor 0.2 \,\, (\phi_d=51^\circ), \quad
d\sim0.6 \lor 0.3 \,\, (\phi_d=129^\circ).
\end{equation}
Since, as we have seen in Subsection~\ref{subsection:conn}, theoretical
estimates prefer $d\sim0.3$, the solutions for $\gamma$ larger than
$90^\circ$ would be favoured. In (\ref{gamma-fact}) and 
(\ref{D-fact-ranges}), we obtain such solutions for both possible values 
of $\phi_d$. 

The contours shown in Fig.~\ref{fig:gam-Add} hold of course also in the 
case of ${\cal A}_{\rm CP}^{\rm dir}(B_d\to\pi^+\pi^-)\sim0$. However, 
we have then to deal with a fourfold discrete ambiguity in the 
extraction of $\gamma$ for each of the two possible values of $\phi_d$. 
Using the input about the cosines of strong phases from factorization, 
$c\sim c'\sim 1$, these fourfold ambiguities are reduced for 
${\cal A}_{\rm CP}^{\rm mix}(B_d\to\pi^+\pi^-)\sim0$ to the twofold ones 
given in (\ref{gamma-fact}). A similar comment applies also to other 
contours in Fig.~\ref{fig:gam-Add}. 

Let us consider the contour corresponding to 
${\cal A}_{\rm CP}^{\rm mix}(B_d\to\pi^+\pi^-)=0.6$, which agrees with 
the central value in (\ref{CP-Bpipi-average}), to discuss this issue in 
more detail. For values of 
$|{\cal A}_{\rm CP}^{\rm dir}(B_d\to\pi^+\pi^-)|\gsim0.5$, we would obtain
no solutions for $\gamma$. If, for instance, 
$|{\cal A}_{\rm CP}^{\rm dir}(B_d\to\pi^+\pi^-)|$ should stabilize at
0.8, we would have an indication for new physics. In the case of
$|{\cal A}_{\rm CP}^{\rm dir}(B_d\to\pi^+\pi^-)|\sim0.5$, the corresponding
horizontal line touches the 
${\cal A}_{\rm CP}^{\rm mix}(B_d\to\pi^+\pi^-)=0.6$ contours, yielding
$\gamma\sim 50^\circ$ and $130^\circ$ for $\phi_d=51^\circ$ and
$\phi_d=129^\circ$, respectively. Moreover, $\theta\sim 90^\circ$ and 
$d\sim 0.4$ would be preferred in this case. For $\theta=\theta'=90^\circ$, 
expression (\ref{H-expr}) implies
\begin{equation}
d=\sqrt{\frac{1-\epsilon^2H}{\xi^2H-1}},
\end{equation}
which yields $d=0.39$ for $H=7.5$. It is amusing to note that 
$\theta=90^\circ$ and $d=0.39$ give for $(\gamma,\phi_d)=
(47^\circ,51^\circ)$ and $(133^\circ,129^\circ)$ the 
observables ${\cal A}_{\rm CP}^{\rm dir}(B_d\to\pi^+\pi^-)=-0.49$
and ${\cal A}_{\rm CP}^{\rm mix}(B_d\to\pi^+\pi^-)=+0.60$, which are both 
in excellent agreement with (\ref{CP-Bpipi-average}). If we reduce the
value of $|{\cal A}_{\rm CP}^{\rm dir}(B_d\to\pi^+\pi^-)|$ below 0.5, 
we obtain a twofold solution for $\gamma$, where the branches on the
left-hand sides correspond to $0^\circ\lsim\theta\lsim 90^\circ$ and
those on the right-hand side to $90^\circ\lsim\theta\lsim 180^\circ$.
Consequently, the latter ones would be closer to factorization, and would
also be in accordance with the PQCD analysis discussed in 
Subsection~\ref{subsection:conn}. As we have seen there, these theoretical
predictions for $d e^{i\theta}$ seem to favour $\gamma>90^\circ$, and 
would hence require that $\phi_d=129^\circ$ in Fig.~\ref{fig:gam-Add}. 
For values of $|{\cal A}_{\rm CP}^{\rm dir}(B_d\to\pi^+\pi^-)|$ below 0.1, 
we would arrive at the fourfold ambiguities for $\gamma$ discussed above. 

It will be very exciting to see in which direction the data will move. 
We hope that the discrepancy between the BaBar and Belle results will
be resolved in the near future.

\begin{figure}
\vspace*{-0.5cm}
$$\hspace*{-1.cm}
\epsfysize=0.2\textheight
\epsfxsize=0.3\textheight
\epsffile{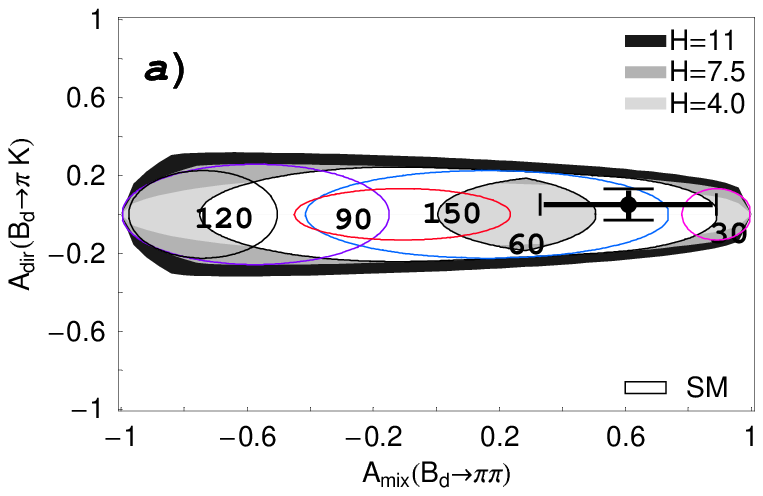} \hspace*{0.3cm}
\epsfysize=0.2\textheight
\epsfxsize=0.3\textheight
 \epsffile{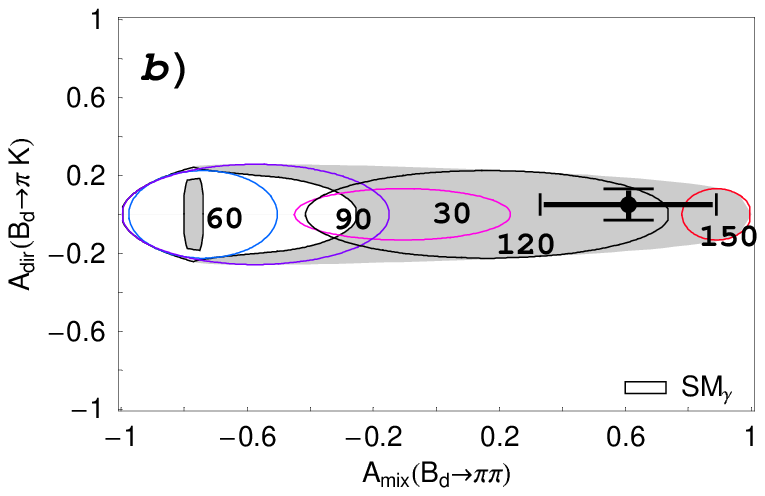}
$$
\vspace*{-0.5cm}
$$\hspace*{-1.cm}
\epsfysize=0.2\textheight
\epsfxsize=0.3\textheight
\epsffile{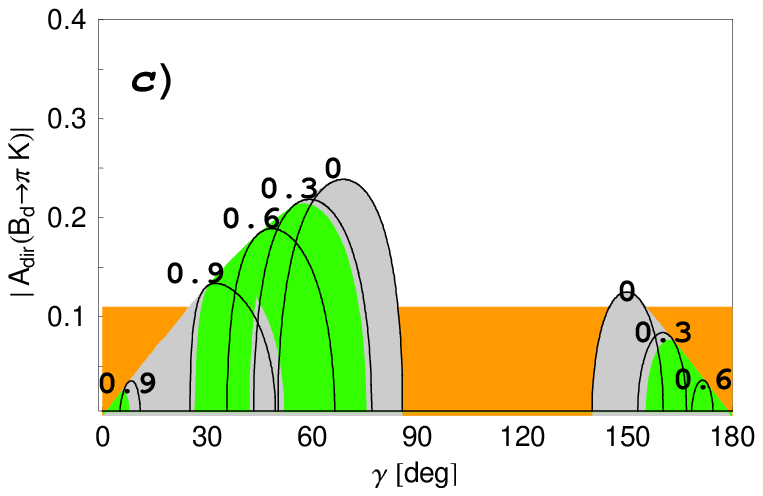} \hspace*{0.3cm}
\epsfysize=0.2\textheight
\epsfxsize=0.3\textheight
 \epsffile{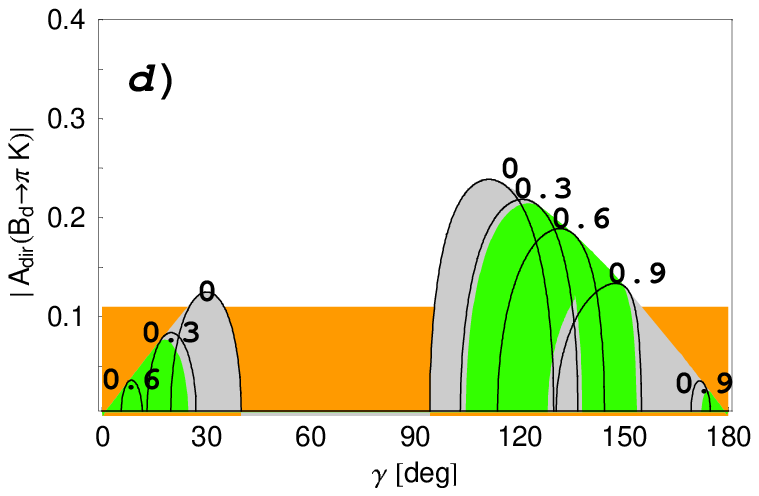}
$$
\vspace*{-0.9cm}
\caption[]{Correlations for the $B_d\to\pi^+\pi^-$, $B_s\to K^+K^-
\approx B_d\to\pi^\mp K^\pm$ system: (a) and (b) are the allowed regions
in the ${\cal A}_{\rm CP}^{\rm mix}(B_d\to\pi^+\pi^-)$--${\cal 
A}_{\rm CP}^{\rm dir}(B_d\to \pi^\mp K^\pm)$ plane for $\phi_d=51^\circ$ 
and $\phi_d=129^\circ$, respectively. In (c) and (d), we consider the 
$\gamma$--$|{\cal A}_{\rm CP}^{\rm dir}(B_d\to \pi^\mp K^\pm)|$ plane for 
various values of ${\cal A}_{\rm CP}^{\rm mix}(B_d\to\pi^+\pi^-)$ in the 
case of $\phi_d=51^\circ$ and $\phi_d=129^\circ$, 
respectively.}\label{fig:Bpipi-BpiK}
\end{figure}

\boldmath
\subsection{Correlations Between $B_d\to\pi^+\pi^-$ and 
$B_d\to\pi^\mp K^\pm$}\label{subsec:Bpipi-BsKK}
\unboldmath
Because of (\ref{CP-rel}) and (\ref{ACP-rep}), it is also interesting to 
consider the CP asymmetry in $B_d\to \pi^\mp K^\pm$ decays instead of
${\cal A}_{\rm CP}^{\rm dir}(B_d\to\pi^+\pi^-)$. The presently available
$B$-factory measurements give
\begin{equation}
{\cal A}_{\rm CP}^{\rm dir}(B_d\to \pi^\mp K^\pm)=\left\{
\begin{array}{ll}
0.04\pm0.16 & \mbox{(CLEO \cite{CLEO-BpiK-CPV})}\\
0.05\pm0.06\pm0.01 &\mbox{(BaBar \cite{BaBar-Bpipi-new})}\\
0.06\pm0.08 & \mbox{(Belle \cite{Belle-Bpipi}),}
\end{array}
\right.
\end{equation}
yielding the average
\begin{equation}\label{ACP-BdpimKp}
{\cal A}_{\rm CP}^{\rm dir}(B_d\to \pi^\mp K^\pm)=0.05\pm0.06.
\end{equation}
On the other hand, inserting the experimental central 
values for ${\cal A}_{\rm CP}^{\rm dir}(B_d\to\pi^+\pi^-)$ and $H$
into (\ref{ACP-rep}) yields
${\cal A}_{\rm CP}^{\rm dir}(B_d\to \pi^\mp K^\pm)\sim0.2$. In view of
the present experimental uncertainties, this cannot be considered as a
discrepancy. If we employ (\ref{CP-rel}) and take into account 
(\ref{xi-def}) and (\ref{Dtheta-def}), we obtain
\begin{displaymath}
\hspace*{-4.3truecm}{\cal A}_{\rm CP}^{\rm dir}(B_d\to \pi^\mp K^\pm)\approx
{\cal A}_{\rm CP}^{\rm dir}(B_s\to K^+K^-)
\end{displaymath}
\begin{equation}
=-\epsilon\xi H\left[
\cos\Delta\theta\pm\frac{\left(u+v d^2\right)
\sin\Delta\theta}{\sqrt{4d^2-\left(u+vd^2\right)^2}}\right]
{\cal A}_{\rm CP}^{\rm dir}(B_d\to\pi^+\pi^-),
\end{equation}
where ${\cal A}_{\rm CP}^{\rm dir}(B_d\to\pi^+\pi^-)$ is given
by (\ref{Adir-expr}), with $d^2$ fixed through (\ref{d2-det}).

In Fig.~\ref{fig:Bpipi-BpiK}, we collect the plots corresponding to
those of the pure $B_d\to\pi^+\pi^-$ correlations given in 
Figs.~\ref{fig:AdAmpipi} and \ref{fig:gam-Add}: in (a) and (b), we 
show the allowed ranges in the ${\cal A}_{\rm CP}^{\rm mix}
(B_d\to\pi^+\pi^-)$--${\cal A}_{\rm CP}^{\rm dir}(B_d\to \pi^\mp K^\pm)$ 
plane for $\phi_d=51^\circ$ and $\phi_d=129^\circ$, respectively, whereas
the curves in (c) and (d) illustrate the corresponding situation in the 
$\gamma$--$|{\cal A}_{\rm CP}^{\rm dir}(B_d\to \pi^\mp K^\pm)|$ plane
for positive values of ${\cal A}_{\rm CP}^{\rm mix}(B_d\to\pi^+\pi^-)$.
We observe that the overlap of the experimental bands gives solutions
for $\gamma$ that are consistent with (\ref{gam-res}), although we
have now also two additional ranges for each $\phi_d$ due to the small 
central value of (\ref{ACP-BdpimKp}).

If we consider the allowed regions in observable space of the direct and
mixing-induced CP asymmetries of the decay $B_s\to K^+K^-$, we obtain a 
very constrained situation. Let us next have a closer look at this 
particularly interesting transition.

\boldmath
\section{Allowed Regions in $B_s\to K^+K^-$ Observable Space}\label{sec:BsKK}
\unboldmath
\subsection{General Formulae}
From a conceptual point of view, the analysis of the decay $B_s\to K^+K^-$
is very similar to the one of $B_d\to\pi^+\pi^-$. If we use (\ref{ACPs-mix}) 
to eliminate $\theta'$ in (\ref{ACPs-dir}), we arrive at 
\begin{equation}\label{Adirs-expr}
{\cal A}_{\rm CP}^{\rm dir}(B_s\to K^+K^-)=
\pm\left[\frac{\sqrt{4\tilde d'^2-\left(u'+v'\tilde d'^2\right)^2}
\sin\gamma}{(1-u'\cos\gamma)+(1-v'\cos\gamma)\tilde d'^2}\right],
\end{equation}
where $u'$ and $v'$ correspond to $u$ and $v$, respectively, and are given by
\begin{equation}\label{us-def}
u'\equiv\frac{{\cal A}_{\rm CP}^{\rm mix}(B_s\to K^+K^-)-
\sin(\phi_s+2\gamma)}{{\cal A}_{\rm CP}^{\rm mix}(B_s\to K^+K^-)\cos\gamma
-\sin(\phi_s+\gamma)}
\end{equation}
\begin{equation}\label{vs-def}
v'\equiv\frac{{\cal A}_{\rm CP}^{\rm mix}(B_s\to K^+K^-)-
\sin\phi_s}{{\cal A}_{\rm CP}^{\rm mix}(B_s\to K^+K^-)\cos\gamma
-\sin(\phi_s+\gamma)}.
\end{equation}
In analogy to (\ref{Adir-expr}), (\ref{Adirs-expr}) is also an exact
expression. Making use of (\ref{U-spin-rel}), the mixing-induced CP 
asymmetry ${\cal A}_{\rm CP}^{\rm mix}(B_s\to K^+K^-)$ allows us to 
eliminate $\theta'$ also in (\ref{H-expr}), thereby providing an 
expression for $\tilde d'^2$. If we take into account, furthermore, 
(\ref{xi-def}) and (\ref{Dtheta-def}), we obtain
\begin{equation}\label{d2-tilde-det}
\tilde d'^2=\frac{A'B'+(2-u'v')S'^2\pm |S'|\sqrt{4A'B'-(A'v'+B'u')^2+
4(1-u'v')S'^2}}{B'^2+v'^2S'^2},
\end{equation}
with
\begin{eqnarray}
A'&\equiv&(\epsilon^2H-1)\xi^2-\epsilon\xi u'(\cos\Delta\theta+\epsilon\xi H)
\cos\gamma\\
B'&\equiv&\epsilon\left[\epsilon\left(1-\xi^2H\right)+\xi v'(\cos\Delta\theta+
\epsilon\xi H)\cos\gamma\right]\\
S'&\equiv&\epsilon\xi\cos\gamma\sin\Delta\theta.
\end{eqnarray}
In the limit of $\Delta\theta=0^\circ$, (\ref{d2-tilde-det}) simplifies to
\begin{equation}
\left.\tilde d'^2\right|_{\Delta\theta=0^\circ}=\frac{A'}{B'}=
\frac{(\epsilon^2H-1)\xi^2-\epsilon\xi u'(1+\epsilon\xi H)
\cos\gamma}{\epsilon\left[\epsilon\left(1-\xi^2H\right)+\xi
v'(1+\epsilon\xi H)\cos\gamma\right]}.
\end{equation}
As in the case of $B_d\to\pi^+\pi^-$, (\ref{Adirs-expr}) and
(\ref{d2-tilde-det}) are invariant under the following symmetry 
transformation:
\begin{equation}\label{sym-rel-s}
\phi_s\to 180^\circ-\phi_s,\quad \gamma\to 180^\circ-\gamma.
\end{equation}
Since $\phi_s$ is negligibly small in the Standard Model, these symmetry
relations may only be of academic interest in the case of $B_s\to K^+K^-$.
On the other hand, $\phi_s$ could in principle also be close to $180^\circ$.
In this case, $B_s\to J/\psi \phi$ would not show CP-violating effects, as 
in the Standard Model. Strategies to distinguish between $\phi_s=0^\circ$
and $180^\circ$ were addressed in \cite{dfn}.

\begin{figure}[t]
$$\hspace*{-1.cm}
\epsfysize=0.2\textheight
\epsfxsize=0.3\textheight
\epsffile{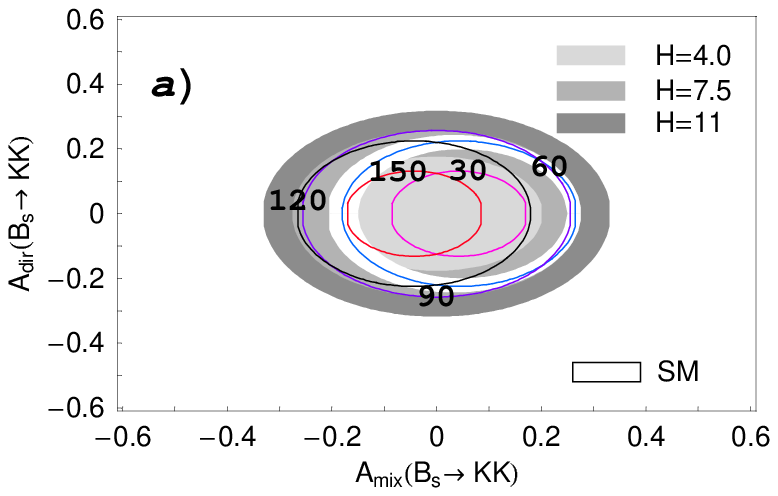} \hspace*{0.3cm}
\epsfysize=0.2\textheight
\epsfxsize=0.3\textheight
 \epsffile{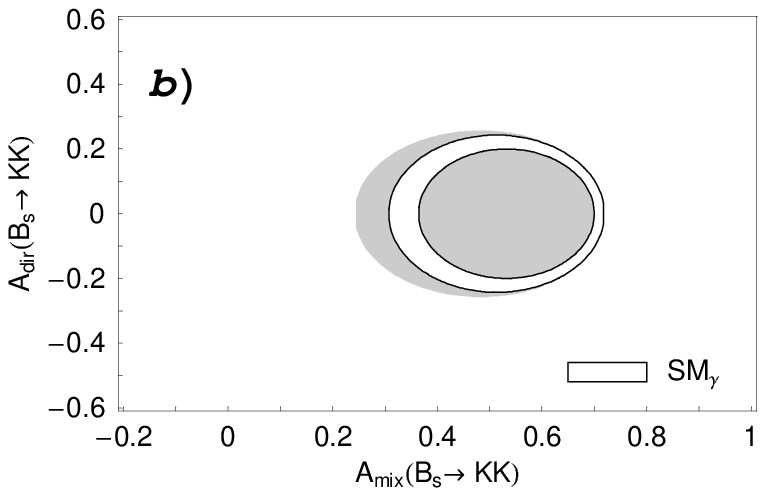}
$$
\vspace*{-0.9cm}
\caption[]{Allowed region in the
${\cal A}_{\rm CP}^{\rm mix}(B_s\to K^+K^-)$--${\cal A}_{\rm CP}^{\rm
dir}(B_s\to K^+K^-)$ plane for (a) $\phi_s=0^\circ$ and various values
of $H$, and (b) $\phi_s=30^\circ$, illustrating the impact of possible
CP-violating new-physics contributions to $B^0_s$--$\overline{B^0_s}$ 
mixing. The SM regions arise if we restrict $\gamma$ to (\ref{gamma-SM}) 
($H=7.5$). We have also included the contours corresponding to various fixed 
values of $\gamma$.}\label{fig:Ams-Ads}
\end{figure}

\begin{figure}
$$\hspace*{-1.cm}
\epsfysize=0.2\textheight
\epsfxsize=0.3\textheight
\epsffile{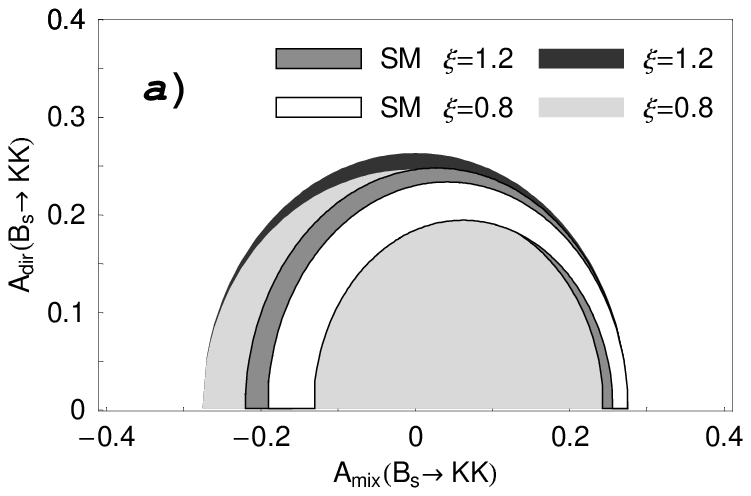} \hspace*{0.3cm}
\epsfysize=0.2\textheight
\epsfxsize=0.3\textheight
\epsffile{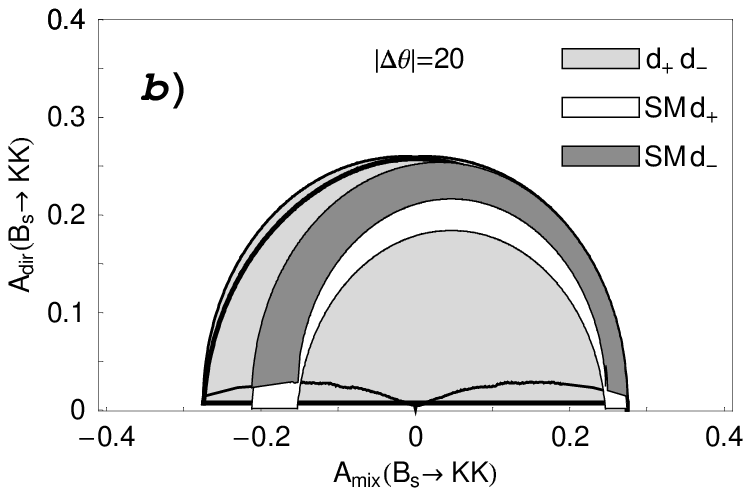}
$$
\vspace*{-0.9cm}
\caption[]{Impact of variations of (a) $\xi\in[0.8,1.2]$, and (b)
$\Delta\theta\in[-20^\circ,+20^\circ]$ on the allowed region in 
the ${\cal A}_{\rm CP}^{\rm mix}(B_s\to K^+K^-)$--${\cal 
A}_{\rm CP}^{\rm dir}(B_s\to K^+K^-)$ plane for $\phi_s=0^\circ$ 
($H=7.5$). In (b), $d_+$ and $d_-$ correspond to the two solutions 
for $\tilde d'^2$ arising in 
(\ref{d2-tilde-det}).}\label{fig:Ams-Ads-xi-Dtheta}
\end{figure}

\subsection{Numerical Analysis}\label{subsec:BsKK-num}
In analogy to our study of $B_d\to\pi^+\pi^-$ in Section~\ref{sec:Bdpipi}, 
we may now straightforwardly calculate the allowed region in the
${\cal A}_{\rm CP}^{\rm mix}(B_s\to K^+K^-)$--${\cal A}_{\rm CP}^{\rm 
dir}(B_s\to K^+K^-)$ plane. In Fig.~\ref{fig:Ams-Ads}, we show these
correlations for (a) a negligible $B^0_s$--$\overline{B^0_s}$
mixing phase $\phi_s$, and (b) a value of $\phi_s=30^\circ$, illustrating 
the impact of possible CP-violating new-physics contributions to 
$B^0_s$--$\overline{B^0_s}$ mixing. We have also indicated the contours 
corresponding to various fixed values of $\gamma$, and the region, which
arises if we restrict $\gamma$ to the Standard-Model range (\ref{gamma-SM}).
In contrast to Fig.~\ref{fig:AdAmpipi}, the allowed region is now
very constrained, thereby providing a narrow target range for run II of
the Tevatron and the experiments of the LHC era. As we have seen in
Subsection~\ref{subsec:Bpipi-BsKK}, the experimental constraints on 
${\cal A}_{\rm CP}^{\rm dir}(B_d\to \pi^\mp K^\pm)$ exclude already
very large direct CP violation in this channel. Because of (\ref{ACP-rep}),
we expect a similar situation in $B_s\to K^+K^-$, which is in accordance
with Fig.~\ref{fig:Ams-Ads}. The allowed range for
${\cal A}_{\rm CP}^{\rm mix}(B_s\to K^+K^-)$ may be shifted significantly 
through sizeable values of $\phi_s$. Such a scenario would be signaled 
independently through large CP-violating effects in the $B_s\to J/\psi\phi$ 
channel, which is very accessible at hadronic $B$ experiments. It is 
interesting to note that if the solution $\phi_d=129^\circ$ should actually 
be the correct one, it would be very likely to have also new-physics effects 
in $B^0_s$--$\overline{B^0_s}$ mixing. If we restrict $\gamma$ 
to the 
Standard-Model range (\ref{gamma-SM}), we even obtain a much more 
constrained allowed region, given by a rather narrow elliptical 
band. 

The sensitivity of the allowed region in the 
${\cal A}_{\rm CP}^{\rm mix}(B_s\to K^+K^-)$--${\cal A}_{\rm CP}^{\rm 
dir}(B_s\to K^+K^-)$ plane on variations of $\xi$ and $\Delta\theta$ 
within reasonable ranges is very small, as can be seen in
Figs.~\ref{fig:Ams-Ads-xi-Dtheta} (a) and (b), respectively. In the latter
figure, we consider $|\Delta\theta|=20^\circ$, and show explicitly the two 
solutions ($d_+$ and $d_-$) for $\tilde d'^2$ arising in
(\ref{d2-tilde-det}). As in Fig.~\ref{fig:Ams-Ads}, we consider again
the whole range for $\gamma$, and its restriction to (\ref{gamma-SM}). 
The shifts with respect to the $\xi=1$, $\Delta\theta=0^\circ$ case are
indeed small, as can be seen by comparing with rescaled 
Fig.~\ref{fig:Ams-Ads} (a). Consequently, the main theoretical 
uncertainty
of our predictions for the $B_s\to K^+K^-$ observable correlations is due to 
the determination of $H$.

It will be very exciting to see whether the measurements at run II of
the Tevatron and at the experiments of the LHC era, where the physics
potential of the $B_s\to K^+K^-$, $B_d\to\pi^+\pi^-$ system can be
fully exploited, will actually hit the very constrained allowed region 
in observable space. In this case, it would be more advantageous not 
to use $H$ for the extraction of $\gamma$, but contours in the 
$\gamma$--$d'$ and $\gamma$--$d$ planes, which can be fixed in a 
theoretically clean way through the CP-violating $B_s\to K^+K^-$ and 
$B_d\to\pi^+\pi^-$ observables, respectively \cite{RF-BsKK}. Making 
then use of $d'=\xi d$, $\gamma$ and the hadronic parameters $d$, $\theta$ 
and $\theta'$ can be determined in a transparent manner. Concerning 
theoretical uncertainties, this is the cleanest way to extract 
information from the $B_s\to K^+K^-$, $B_d\to\pi^+\pi^-$ system. In 
particular, it does not rely on (\ref{U-fact}). It should be noted that
this approach would also work, if $\phi_s$ turned out to be sizeable. 
This phase could then be determined through $B_s\to J/\psi\phi$
\cite{ddfn,dfn}.

\boldmath
\subsection{Comments on Factorization}
\unboldmath
Using the same input from factorization as in Subsection~\ref{subsec:fact},
we obtain the following simplified expressions for the contours in the 
$\gamma$--$d'$ and $\gamma$--$d$ planes:
\begin{equation}\label{g-d-fact}
d'=\epsilon\left(\frac{c'\pm\sqrt{c'^2-u'v'}}{v'}\right),\quad
d=\frac{-c\pm\sqrt{c^2-uv}}{v},
\end{equation}
where $u'$, $v'$ and $u$, $v$ are given in (\ref{us-def}), (\ref{vs-def})
and (\ref{u-def}), (\ref{v-def}), respectively, and $c'$ and $c$ are
defined in (\ref{c-cp-def}). On the other hand, the general expressions 
derived in \cite{RF-BsKK} that do not rely on factorization simplify 
for vanishing direct CP asymmetries in $B_s\to K^+K^-$ and $B_d\to\pi^+\pi^-$ 
as follows:
\begin{equation}\label{g-d-exact}
d'=\epsilon\left|\frac{1\pm\sqrt{1-u'v'}}{v'}\right|,\quad
d=\left|\frac{-1\pm\sqrt{1-uv}}{v}\right|.
\end{equation}
Consequently, since $d'$ and $d$ are by definition positive parameters,
the input from factorization would allow us to reduce the number of discrete 
ambiguities in this case. We have encountered a similar feature in our 
discussion of $B_d\to\pi^+\pi^-$ in Subsection~\ref{subsec:fact}. As noted
in \cite{RF-BsKK}, in order to reduce the number of discrete ambiguities,
also the contours in the $\gamma$--$d$ and $\gamma$--$d'$ plane
specified through (\ref{d2-det}) and (\ref{d2-tilde-det}), respectively, 
are very helpful.

\begin{figure}
\vspace*{-0.2cm}
$$\hspace*{-1.cm}
\epsfysize=0.2\textheight
\epsfxsize=0.3\textheight
\epsffile{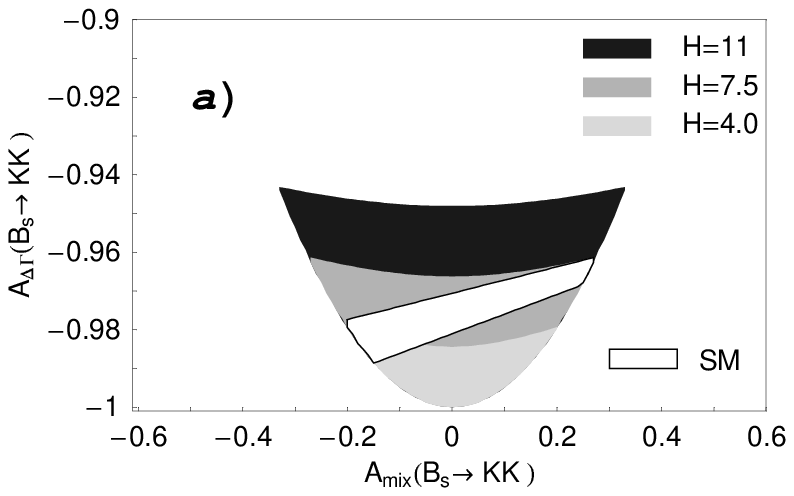} \hspace*{0.3cm}
\epsfysize=0.2\textheight
\epsfxsize=0.3\textheight
\epsffile{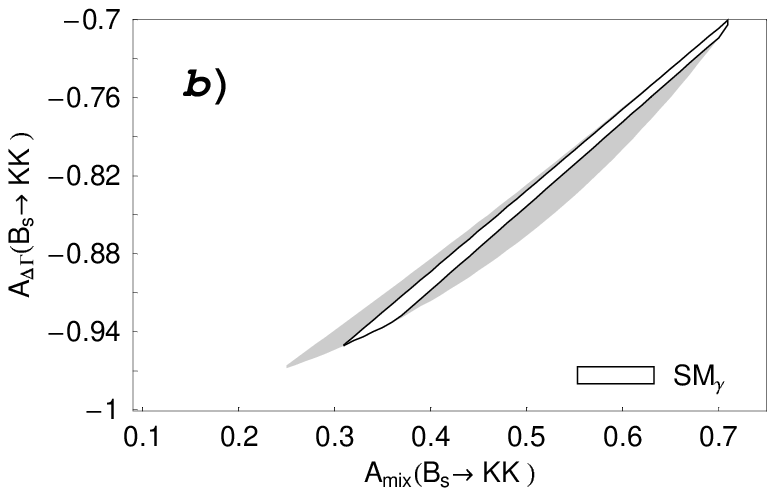}
$$
\vspace*{-0.9cm}
\caption[]{Allowed region in the
${\cal A}_{\rm CP}^{\rm mix}(B_s\to K^+K^-)$--${\cal A}_{\Delta\Gamma}
(B_s\to K^+K^-)$ plane for (a) $\phi_s=0^\circ$ and various values
of $H$, and (b) $\phi_s=30^\circ$, illustrating the impact of 
CP-violating new-physics contributions to $B^0_s$--$\overline{B^0_s}$ 
mixing. The SM regions arise if we restrict $\gamma$ to (\ref{gamma-SM}) 
($H=7.5$).}\label{fig:Ams-ADG}
\end{figure}

\boldmath
\subsection{The ${\cal A}_{\rm CP}^{\rm mix}(B_s\to 
K^+K^-)$--${\cal A}_{\Delta\Gamma}(B_s\to K^+K^-)$ Plane}
\unboldmath
Let us finally consider the observable
${\cal A}_{\Delta\Gamma}(B_s\to K^+K^-)$ appearing in (\ref{time-CP-asym}),
which may be accessible due to a sizeable width difference $\Delta\Gamma_s$
of the $B_s$ system. Interestingly, this quantity may also be extracted from
the ``untagged'' rate
\begin{equation}
\Gamma(B^0_s(t)\to K^+K^-)+\Gamma(\overline{B^0_s}(t)\to K^+K^-)
\propto R_{\rm H}e^{-\Gamma_{\rm H}^{(s)}t}+
R_{\rm L}e^{-\Gamma_{\rm L}^{(s)}t}
\end{equation}
through
\begin{equation}
{\cal A}_{\Delta\Gamma}(B_s\to K^+K^-)=
\frac{R_{\rm H}-R_{\rm L}}{R_{\rm H}+R_{\rm L}},
\end{equation}
where $\Delta\Gamma_s\equiv \Gamma_{\rm H}^{(s)}-\Gamma_{\rm L}^{(s)}$
is negative in the Standard Model. Using parametrization (\ref{Bs-ampl}),
we obtain \cite{RF-BsKK}
\begin{equation}
{\cal A}_{\Delta\Gamma}(B_s\to K^+K^-)=
-\left[\frac{\cos(\phi_s+2\gamma)+2\tilde d'\cos\theta'\cos(\phi_s+\gamma)+
\tilde d'^2\cos\phi_s}{1+2\tilde d'\cos\theta'\cos\gamma+
\tilde d'^2}\right].
\end{equation}
An important difference with respect to the direct CP asymmetry
(\ref{ACPs-dir}) is that -- as in (\ref{ACPs-mix}) -- only $\cos\theta'$
terms appear in this expression. Consequently, using the mixing-induced
CP asymmetry ${\cal A}_{\rm CP}^{\rm mix}(B_s\to K^+K^-)$ to eliminate
$\cos\theta'$, we arrive at
\begin{equation}\label{ADG-expr-calc}
{\cal A}_{\Delta\Gamma}(B_s\to K^+K^-)=-\left[
\frac{\cos(\phi_s+2\gamma)-u'\cos(\phi_s+\gamma)+\left\{\cos\phi_s-v'
\cos(\phi_s+\gamma)\right\}\tilde d'^2}{\left(1-u'\cos\gamma\right)+
\left(1-v'\cos\gamma\right)\tilde d'^2}\right].
\end{equation}
In contrast to (\ref{Adirs-expr}), no sign ambiguity appears in this 
expression; in the former, it is due to 
$\sin\theta'=\pm\sqrt{1-\cos^2\theta'}$. The square root in 
(\ref{Adirs-expr}) ensures that
\begin{equation}
|\cos\theta'|=\frac{|u'+v'\tilde d'^2|}{2\tilde d'}\leq1.
\end{equation}
If we fix $\tilde d'$ through (\ref{d2-tilde-det}) and insert it into
(\ref{ADG-expr-calc}), we have to require, in addition, that this relation
is satisfied. We may then perform an analysis similar to the one for the 
observables ${\cal A}_{\rm CP}^{\rm mix}(B_s\to K^+K^-)$ and 
${\cal A}_{\rm CP}^{\rm dir}(B_s\to K^+K^-)$ given above. 

In Fig.~\ref{fig:Ams-ADG}, we show the allowed region in the 
${\cal A}_{\rm CP}^{\rm mix}(B_s\to K^+K^-)$--${\cal A}_{\Delta\Gamma}
(B_s\to K^+K^-)$ plane for (a) the Standard-Model case of $\phi_s=0^\circ$,
and (b) a value of $\phi_s=30^\circ$, illustrating the impact of possible
new-physics contributions to $B^0_s$--$\overline{B^0_s}$ mixing. It should
be noted that the width difference $\Delta\Gamma_s$ would be modified
in the latter case as follows \cite{dfn,grossman}:
\begin{equation}\label{DG-NP}
\Delta\Gamma_s=\Delta\Gamma_s^{\rm SM}\cos\phi_s.
\end{equation}
As in Fig.~\ref{fig:Ams-Ads}, we have also included the regions, which 
arise if we restrict $\gamma$ to (\ref{gamma-SM}). We observe that
${\cal A}_{\Delta\Gamma}(B_s\to K^+K^-)$ is highly constrained within 
the Standard Model, yielding
\begin{equation}
-1\lsim {\cal A}_{\Delta\Gamma}(B_s\to K^+K^-) \lsim -0.95.
\end{equation}
Moreover, it becomes evident that this observable may be affected 
significantly through sizeable values of $\phi_s$. Unfortunately, 
the width difference $|\Delta\Gamma_s|$ would be reduced in this case 
because of (\ref{DG-NP}), thereby making measurements relying on a 
sizeable value of this quantity more difficult.

\section{Conclusions and Outlook}\label{sec:concl}
In our paper, we have used recent experimental data to analyse allowed 
regions in the space of CP-violating $B\to\pi K$,  $B_d\to\pi^+\pi^-$ 
and $B_s\to K^+K^-$ observables that arise within the Standard Model. The 
main results can be summarized as follows:
\begin{itemize}
\item As far as $B\to\pi K$ decays are concerned, the combinations of
charged and neutral modes appear to be most exciting. We have presented
contour plots, allowing us to read off the preferred ranges for $\gamma$
and strong phases $\delta_{\rm c,n}$ directly from the experimental data.
The charged and neutral $B\to\pi K$ decays point both towards 
$\gamma>90^\circ$. On the other hand, they prefer $|\delta_{\rm c}|$ to be
smaller than $90^\circ$, and $|\delta_{\rm n}|$ to be larger than $90^\circ$. 
This puzzling situation, which was also pointed out in \cite{BF-neutral2}, 
may be an indication of new-physics contributions to the electroweak penguin 
sector, but the uncertainties are still too large to draw definite 
conclusions. It should be kept in mind that we may also have ``anomalously''
large flavour-symmetry breaking effects.
\item The present data on the CP-averaged $B_d\to\pi^\mp K^\pm$ and 
$B_d\to\pi^+\pi^-$ branching ratios allow us to obtain rather strong
constraints on the penguin parameter $d e^{i\theta}$. A comparison of the
experimental curves with the most recent theoretical predictions for this
parameter is not in favour of an interpretation within the 
Standard Model;
comfortable agreement between theory and experiment could be achieved for 
values of $\gamma$ being larger than $90^\circ$.
\item The constraints on $d e^{i\theta}$ have interesting implications
for the allowed region in the space of the mixing-induced and direct CP 
asymmetries of the decay $B_d\to\pi^+\pi^-$. Taking into account the
first measurements of these observables at the $B$ factories, we arrive
at the following picture:
\begin{itemize}
\item For the $B^0_d$--$\overline{B^0_d}$ mixing phase $\phi_d=51^\circ$, 
the data favour a value of $\gamma\sim 50^\circ$. In this case, 
$\phi_d=2\beta$ and $\gamma$ would both agree with the results of the 
usual indirect fits of the unitarity triangle. 
\item For the $B^0_d$--$\overline{B^0_d}$ mixing phase 
$\phi_d=180^\circ-51^\circ=129^\circ$, the data favour a value of 
$\gamma\sim 130^\circ$, i.e.\ larger than $90^\circ$. In this case, 
$\phi_d$ would require CP-violating new-physics contributions to 
$B^0_d$--$\overline{B^0_d}$ mixing, so that also the results of the usual 
indirect fits of the unitarity triangle for $\gamma$ may no longer hold.
\end{itemize}
As we have noted above, $\gamma$ may actually be larger than $90^\circ$,
which would then require the unconventional solution $\phi_d=129^\circ$.
Consequently, it is very important to resolve the twofold ambiguity arising 
in the extraction of $\phi_d$ from ${\cal A}_{\rm CP}^{\rm mix}(B_d\to J/\psi
K_{\rm S})=-\sin\phi_d$ directly through a measurement of the sign of
$\cos\phi_d$. 
\item We have provided the formalism to take into account the parameters
$\xi$ and $\Delta\theta$, affecting the theorectical accuracy of our
approach, in an exact manner, and have studied their impact in detail.
\item In the case of the decay $B_s\to K^+K^-$, we obtain a very constrained
allowed region in the space of the corresponding CP-violating observables, 
thereby providing a narrow target range for run II of the Tevatron and the 
experiments of the LHC era. Here the impact of variations of $\xi$ and
$\Delta\theta$ within reasonable ranges is practically negligible. 
On the basis of the present data on direct CP 
violation in $B_d\to\pi^\mp K^\pm$, we do not expect a very 
large value of 
${\cal A}_{\rm CP}^{\rm dir}(B_s\to K^+K^-)$, which is also in accordance 
with the allowed range derived in this paper. On the other
hand, CP-violating new-physics contributions to $B^0_s$--$\overline{B^0_s}$ 
mixing may shift the range for ${\cal A}_{\rm CP}^{\rm mix}(B_s\to K^+K^-)$
significantly. 
\item Using a moderate input from factorization about the cosines of
CP-conserving strong phases, our analysis could be simplified, and the 
number of discrete ambiguities arising in the extraction of $\gamma$
could be reduced. 
\end{itemize}
It will be very exciting to see in which direction the experimental results
for the $B\to\pi K$,  $B_d\to\pi^+\pi^-$ 
and $B_s\to K^+K^-$ observables will move. Unfortunately, the present 
measurements of the CP asymmetries in $B_d\to\pi^+\pi^-$ by BaBar and 
Belle are not fully consistent with each other. We hope that this 
discrepancy will be resolved soon. As we have pointed out in our analysis, 
we may obtain valuable insights into CP violation and the world of penguins 
from such measurements. A first analysis of $B_s\to K^+K^-$ will already be 
available at run II of the Tevatron, where $B^0_s$--$\overline{B^0_s}$ 
mixing should also be discovered, and $B_s\to J/\psi \phi$ may indicate a 
sizeable value of $\phi_s$. At the experiments of the LHC era, in particular 
LHCb and BTeV, the physics potential of the $B_s\to K^+K^-$, 
$B_d\to \pi^+\pi^-$ system to explore CP violation can then be fully 
exploited.

\section*{Acknowledgements}
R.F. is grateful to Wulfrin Bartel and Masashi Hazumi for discussions and
correspondence on the recent Belle measurement of $B_d\to\pi^+\pi^-$, in 
particular the employed sign conventions, and would like to thank the 
Theoretical High-Energy Physics Group of the Universitat Aut\`onoma de 
Barcelona for the kind hospitality during his visit. J.M. acknowledges 
financial support by CICYT Research Project AEN99-07666 and by Ministerio de 
Ciencia y Tecnologia from Spain.

\end{document}